\documentclass[twocolumn,floatfix,pre,aps,a4paper,showpacs]{revtex4-1}
\pdfoutput=1
\usepackage{graphicx}
\usepackage{amssymb}
\usepackage{amsmath}
\usepackage[utf8]{inputenc}
\usepackage{cancel}
\usepackage[usenames,dvipsnames]{color}
\usepackage{psfrag}
\usepackage{epsfig}
\usepackage{bbm}
\usepackage{bm}
\usepackage{dsfont}
\usepackage{soul}
\usepackage{colonequals}
\usepackage{float}
\usepackage{enumitem}
\begin{document}
\title{Force-linearization closure for non-Markovian Langevin systems with time delay} 
\author{Sarah A.~M.~Loos and Sabine H.~L.~Klapp}
\affiliation{
  Institut f\"ur Theoretische Physik,
  Hardenbergstr.~36,
  Technische Universit\"at Berlin,
  D-10623 Berlin,
  Germany}
\date{\today}
\begin{abstract}
This paper is concerned with the Fokker-Planck (FP) description of classical stochastic systems with discrete time delay. The non-Markovian character of the corresponding Langevin dynamics naturally leads to a coupled infinite hierarchy of FP equations for the various $n$-time joint distribution functions. Here we present a novel approach to close the hierarchy at the one-time level based on a linearization of the deterministic forces in all members of the hierarchy starting from the second one. This leads to a closed equation for the one-time probability density in the steady state. Considering two generic nonlinear systems, a colloidal particle in a sinusoidal or bistable potential supplemented by a linear delay force, we demonstrate that our approach yields a very accurate representation of the density as compared to quasi-exact numerical results from direct solution of the Langevin equation. Moreover, the results are significantly improved against those from a small-delay approximation and a perturbation-theoretical approach. We also discuss the possibility of accessing transport-related quantities, such as escape times, based on an additional Kramers approximation. Our approach applies to a wide class of models with nonlinear deterministic forces. 
 \end{abstract}
\maketitle
\section{Introduction \label{SEC:INTRO}}
Many (non-)equilibrium systems from the macroscopic world down to the quantum level are governed by dynamical equations which involve both, noise and delay in time \cite{Longtin2010, Schoell2016,Schoell2008}. Noise due to imperfections or environmental influences is essentially omnipresent in many real-world systems and experimental setups (e.g., at the cantilever of an Atomic Force Microscope \cite{Montinaro2012}). On the mathematical level it often also results from the presence of hidden degrees of freedom, which have disappeared in the context of a coarse-graining procedure. Similarly, time delay can have various sources. One example are finite information processing and reaction times, which occur for example in social systems (e.g., financial markets \cite{Stoica2004} or economic processes \cite{Voss2002}). Such latencies as well appear in neural systems due to finite signal transmission and refractory times (which, e.g., become apparent in the human pupil reflex \cite{Longtin1990}, or in stick balancing experiments \cite{Milton2013,*Milton2009a}). In biological systems, delays can be caused by maturation times (as in population growth \cite{Goel1971} or prey-predator dynamics \cite{Das2012}), or, on the biomolecular level, by biochemical reactions kinetics (such as the transcriptional and translational delay in gene networks \cite{Parmar2015, *Bratsun2005, *Gupta2013}). In laser dynamics with optical feedback, delay is induced by the traveling time of laser light on its round trip in the cavity \cite{Masoller2002}. On an even smaller scale, time delays occur in quantum-optical systems coupled to a structured photonic reservoir \cite{Hein2015, *Lu2017}.
\par
Beyond these intrinsic delays, time delay is an important issue in experimental setups with feedback control, where the delay is due to the finite time to proceed the output (plus the signal transmission times to obtain the information and feed it back into the system). Indeed, feedback control has emerged as an important tool to manipulate small systems. Important representatives for such systems are colloids in a thermal bath under time delayed feedback control \cite{Braun2013, *Braun2015,*Haeufle2016,Lichtner2012,*Lichtner2010, Hennig2009}. All these examples involve noise and delay, illustrating that due to the generality and omnipresence of such features, noisy systems with time delays occur on all scales. In many of these cases, the interplay of noise and delay moreover leads to intriguing dynamical behavior such as multistability and stochastic switches \cite{Masoller2002, Masoller2003}.
\par
However, even in the classical case the mathematical description of noisy systems with time delay continues to be a major challenge. In presence of a discrete time delay $\tau$, the standard mathematical equation for stochastic motion, that is, the (under- or overdamped) Langevin equation (LE), becomes a stochastic {\em delay} differential equation (SDDE), or, in a more physical language, a non-Markovian Langevin equation. The usual concepts and solutions for Markovian LEs then do not apply and indeed, the development of a general solution method for SDDEs is yet an unresolved problem \cite{Rosinberg2015}. Moreover, contrary to the Markovian systems, the route from the LE to a Fokker-Planck equation (FPE) for the corresponding probability density is more involved \cite{Rosinberg2015,Guillouzic1999,Frank2005,*Frank2005a,Giuggioli2016}. The time delay alone leads to an infinite hierarchy of coupled FP equations for the $n$-time (joint) probability densities \cite{Rosinberg2015,Guillouzic1999,Frank2005,*Frank2005a,Giuggioli2016}, for which as well no general solution has been found to date. Similar hierarchy problems also occur in other contexts in statistical physics. A well known example is the Bogoliubov–Born–Green–Kirkwood–Yvon (BBGKY) hierarchy, in which the time evolution of the one-particle density depends upon the two-particle density and so on. For this problem, various closing strategies have been proposed, such as the simple mean-field (factorization) approximation and the more sophisticated dynamical density functional theory (involving an adiabatic approximation) \cite{Marconi1999, *Menzel2016}. Another closure example is the mode coupling theory for glassy systems \cite{Das2004, *Janssen2015}. We stress, however, that in all these examples the hierarchical structure emerges due to (conservative) particle-particle interactions. In contrast, the hierarchy of FPEs appearing in time-delayed systems -- although having a similar structure -- 
arises due to delay-induced \textit{temporal} correlations.
\par
These aspects are also of major relevance in the emerging field of stochastic thermodynamics \cite{Seifert2012,Sagawa2012,*Abreu2012,*Barato2014,*Mandal2012,*Parrondo2015,*Kutvonen2016, Loos2014}. Indeed, concepts like entropy production, fluctuation theorems and information exchange have been widely formulated (and tested experimentally \cite{Kim2007, Koski2014}) for Markovian noisy systems, for which the correspondence between the various levels of description (LE, FPE, path integrals) can readily be utilized. For example, the entropy production from the system can be directly calculated via the probability density and the corresponding probability current \cite{Seifert2012}. For non-Markovian systems, all these concepts have to be revisited (see \cite{Rosinberg2015,Jiang2011,*Munakata2009,*Munakata2014} for recent work in this direction).
\par
The only solvable class of delayed noisy systems are those with {\em linear} forces, where all the $n$-time probability densities are given by multivariate Gaussian distributions. Here, the system can be solved exactly on the LE \cite{Kuechler1992} and the FPE level \cite{Giuggioli2016}, or by utilizing both levels of description \cite{Frank2001}. In the more general case of nonlinear systems, one has to rely on approximations \cite{Longtin2010}. For continuous systems, the two most established strategies are a Taylor expansion of the LE in $\tau$ \cite{Guillouzic1999,Guillouzic2000} and a perturbation theory on the level of the FPE \cite{Frank2005,*Frank2005a}, where the entire delay force is treated as a small deviation from the Markovian dynamics. 
\par
 In this work, we propose a novel FPE-based approach, which we call ``force linearization closure'' (FLC). The main idea is to linearize the deterministic force in all members of the FPE hierarchy starting from the second one. We then utilize the (Gaussian) solution for the linear, delayed higher-order FPEs to obtain a closure of the FPE for the one-time probability density in the nonlinear case. Our strategy can be applied whenever the linearized system has a stable steady state.

 \par
 We apply this concept to a classical system, namely, an (overdamped) colloidal particle subject to a nonlinear static force and a time-delayed force representing a feedback control. The control ``target'' is the particle position which, due to the typical (micron-scale) size of a colloid, is readily accessible in real space experiments and simulations \cite{Braun2013, *Braun2015,*Haeufle2016,Lichtner2012,*Lichtner2010, Hennig2009}. Feedback control can be implemented, e.g., as a co-moving ``optical tweezer'' \cite{Loos2014, Kotar2010, Kim2007,Qian2013,*Balijepalli2012, Gernert2014, Gernert2015}, which can be well approximated by a co-moving quadratic potential \cite{Woerdemann2012, Kotar2010, Kim2007}. Depending on the experimental setup, a time delay naturally arises during the position measurement and the adjustment of the tweezer, or it can be intentionally implemented as a feature. As a result, the system is subject to a {\em linear} time-delayed feedback force.
\par
Regarding the nonlinear static force (or potential, respectively) we consider two paradigmatic examples: first, a periodic sinusoidal (``washboard") potential, which is frequently used to model diffusion and transport on rough surfaces \cite{Reimann2001,Juniper2016,Emary2012} and can be realized experimentally e.g. by optical landscapes \cite{Juniper2016}. The impact of time delay on diffusion in washboard potentials has already been studied on the basis of a heuristic delayed FPE \cite{Loos2014, Gernert2014, Gernert2015} and by simulations \cite{Hennig2009}. Our second example is a colloidal particle in a double-well potential. Bistable noisy systems often serve to study escape problems \cite{Kramers1940, Hanggi1990, Risken1984, Gardiner2002}. Bistable systems with delay have been studied theoretically, e.g., regarding the Kramers rate \cite{Guillouzic2000, Goulding2007, Masoller2003} and in the context of coherence resonance (see the study of Tsimring and Pikovsky \cite{Tsimring2001}, who provided a solution based on a discretization procedure, and subsequent work \cite{Tsimring2001, Masoller2003, Du2015, Xiao2016}). Experimental realizations have been suggested in Refs.~\cite{Goulding2007, Piwonski2005}, which involve the polarization dynamics of lasers with optical feedback (where $\tau$ is associated with the external cavity length).
\par
The remainder of this paper is organized as follows. We start by briefly reviewing the Langevin and the Fokker-Planck description of noisy overdamped systems with discrete time delay, as well as the two main earlier approximations for the one-time probability density $\rho_1$ (Sec.\,\ref{Theory}). We then introduce the force linearization closure and provide the main steps to derive its key equation, a closed FPE for $\rho_1$ (Sec.\,\ref{SEC:FLC}). In the second part of the paper, we apply the FLC to the two examples mentioned above and compare the results with quasi-exact numerical results obtained from Brownian dynamics (BD) simulations, i.e., direct simulations of the delayed Langevin equation. We demonstrate that the FLC renders a very good approximation of the one-time probability density, which is, moreover, more accurate than the predictions from the two main earlier approaches. Furthermore, we briefly discuss the possibility to estimate escape times (which are rigorously connected to two-time probability densities) within the novel approach (Sec.\,\ref{SEC:Escape-times}). We summarize and conclude in Sec.\,\ref{SEC:Conclusion}. The paper contains four appendices with additional theoretical results (e.g., the second member of the FPE hierarchy), and technical aspects of the calculations.
\section{Theoretical framework \label{Theory}}
\subsection{Delayed Langevin and Fokker-Planck equation \label{model}}
We consider an overdamped Brownian particle at the time-dependent stochastic position $\chi(t)$ which represents the dynamical quantity of interest. The particle moves in a static external potential $V_\mathrm{s}(x)$, with spatial variable $x\!\in\Omega$, where $\Omega$ denotes the spatial domain of the system. The particle is further subject to a ``delay force'' $F_{\mathrm{d}}$, which depends on the instantaneous and on the \textit{delayed} particles position, i.e., at a given time $t$ on $\chi(t)$ and $\chi(t-\tau)$, with $\tau\!>\!0$ being the delay time. The total deterministic force is hence
\begin{equation}\label{EQ:F}
F(x,x_\tau)\!=\!F_\mathrm{s}(x)+F_\mathrm{d}(x,x_\tau),
\end{equation}
{with $x_\tau \!\in\Omega$ being a second spatial variable needed due to the two involved particle positions in $F_\mathrm{d}$}, and $F_\mathrm{s}(x)\!=\!-\partial_x V_\mathrm{s}(x)$ being the negative of the spatial derivative of $V_\mathrm{s}$. The delayed Langevin equation (LE) reads
\begin{equation}\label{EQ:LE}
{\mathrm{d} {\chi}({t}) } /{\mathrm{d}{t}} = {\gamma^{-1}}F\left[ \chi({t}), \chi({t}-{\tau})\right]+  \sqrt{2 {D_0}}\, {\Gamma}({t}),
\end{equation} 
where $\gamma$ is the friction coefficient of the surrounding medium at temperature ${\cal T}$ and $\Gamma(t)$ denotes the Langevin force. The latter introduces additive Gaussian white noise to the system, i.e., $\langle \Gamma(t) \rangle \!=\! 0 ~\forall t$ and $\langle \Gamma(t)\Gamma(t') \rangle \!=\!  \delta(t-t')~ \forall t,t'$, with $\langle ... \rangle$ denoting the ensemble average and $\delta(t)$ the Delta distribution. Further, $D_0$ is the short-time diffusion coefficient satisfying $D_0\!=\!k_{\mathrm{B}}{\cal T}/\gamma$ (with $k_{\mathrm{B}}$ being the Boltzmann constant) \cite{Risken1984}. We measure the time in units of the ``Brownian'' time scale $\tau_\mathrm{B}\!=\!\sigma^2/D_0$, where $\sigma$ is the particle diameter. This is the time in which a free Brownian particle (no external forces) travels over a distance equal to its own size.\par
As usual for delay equations, the evolution of the dynamical variable (here $\chi$) depends on a given history function $\phi$, which serves as an initial condition $\chi(t)\!=\!\phi(t), ~\forall t \in [-\tau,0]$.\par
We emphasize that the theoretical framework discussed in this paper does not just apply to Brownian particles, but to other natural and artificial systems as well. Having this in mind, the friction and diffusion coefficient, the Brownian time scale and the particle diameter introduced before just provide energy, time and spatial length scales, which one can adjust to the system under consideration.\par
As firstly shown in \cite{Guillouzic1999}, it is possible to derive from the delayed LE a Fokker-Planck equation (FPE) for the (one-time) probability density $\rho_1(x,t)\!=\!\langle \delta(x-\chi(t)) \rangle $ in a similar manner to the Markovian case. We particularly refer to Ref.\,\cite{Frank2005,*Frank2005a}, which presents a derivation based upon Novikov's theorem \cite{Novikov1965}. Novikov's theorem generally links the variational derivative of an arbitrary functional of the Langevin force, $\Lambda[\Gamma]$, to the correlations between that functional and the Langevin force. For Gaussian white noise it reads \cite{Novikov1965, Frank2005}:
 \begin{equation}\label{EQ:Novikov}
\big \langle \Lambda[\Gamma]\Gamma(t)  \big \rangle = \Bigg \langle \frac{\delta \Lambda[\Gamma]}{\delta \Gamma(t)}\Bigg\rangle.
\end{equation}
{A central step in the derivation in \cite{Frank2005} is the usage of Novikov's theorem to express the correlations 
$\big \langle \left\{ \delta(x-\chi(t))\delta(x_\tau-\chi(t-\tau))\right\}\Gamma(t)  \big \rangle$ by the related variational derivatives.}
For the case of additive noise, the delayed FPE for $\rho_1$ is given by \cite{Guillouzic1999, Frank2001, Frank2005} 
\begin{subequations}\label{EQ:dFPE}
\begin{align}\label{EQ:dFPEa}
{\partial_t} \rho_1(x,t)=& - {\gamma^{-1}}{\partial_x}\!\left[\widehat{F}(x,t)\rho_1(x,t) \right] \nonumber \\
&+ D_0 {\partial_{xx}} \rho_1(x,t) ,
\end{align}
where
\begin{align}\label{EQ:dFPEb}
\widehat{F}(x,t)=&F_\mathrm{s}(x)+\!\int_{\Omega}\!\!  F_\mathrm{d}(x,x_\tau\!)\rho_{\mathrm{c}}( x_{\tau}, t\!-\!\tau | x,t) \mathrm{d}x_{\tau} .
\end{align}
\end{subequations}
For the sake of a shorter notation, we drop here and in the following the dependency on the history function, i.e., $\rho_1(x,t)\equiv\rho_1(x,t|\phi)$ and $\rho_{\mathrm{c}}( x_{\tau}, t\!-\!\tau | x,t)\equiv\rho_{\mathrm{c}}( x_{\tau}, t\!-\!\tau | x,t;\phi)$. Comparing Eq.\,(\ref{EQ:dFPEa}) to an ordinary FPE for Markovian systems, one notes the appearance of a delay-averaged drift term $\widehat{F}$ instead of the usual drift term, which only depends on instantaneous quantities. Specifically, the delay-averaged drift involves the conditional probability $\rho_{\mathrm{c}}$, which is related to the two-time (joint) probability density via $\rho_2( x, t ;x_{\tau},t\!-\!\tau)\!=\!\rho_{\mathrm{c}}(x_{\tau},t\!-\!\tau | x, t)\rho_1( x, t)$ [with $\int_{\Omega} \rho_2( x, t ;x_{\tau},t-\!\tau) \mathrm{d}x_\tau\! =\!\rho_{\mathrm{1}}( x,t)$]. Equation (\ref{EQ:dFPE}) is hence not self-sufficient. The delayed FPE for $\rho_2$, on the other hand, involves $\rho_3$ (as explicitly shown in the Appendix\,\ref{APP:FPE_rho2}), and so forth. Thus, an infinite hierarchy of equations emerges, whose $n\!+\!1$st member depends on the $n$-time probability density $\rho_n(x,t;...;x_{n\tau},t-n\tau)$. By finding suitable approximations for $\rho_\mathrm{c}$, Eqs.\,(\ref{EQ:dFPE}) can be closed and the hierarchy truncated. 
This is our objective in the present work.
\subsection{Earlier approaches to approximate the one-time probability density}
One such approach, involving a first-order perturbation-theoretical (PT) ansatz for the density on level of the FPE, was previously introduced in Ref.\,\cite{Frank2005}. Within PT, the delay force $F_\mathrm{d}$ is regarded as a {small perturbation} to the non-delayed dynamics, i.e., $|F_{\mathrm{d}}|\!\ll\!|F_{\mathrm{s}}|$. 
The resulting first-order equation has the same form as Eq.\,(\ref{EQ:dFPEa}), but the $\widehat{F}$-term [Eq.\,(\ref{EQ:dFPEb})] contains the conditional probability density with respect to the {unperturbed}, i.e., non-delayed, system, which we will refer to as $\rho^{F_d\equiv0}_{c}$. The latter follows a closed (Markovian) FPE, and consequentially, the combination of the two {FPEs is self-sufficient}. From a practical perspective, the PT approach is particularly appealing for systems where an analytical expression for $\rho^{F_d\equiv0}_\mathrm{c}$ is available. However, this is the case only for very few nonlinear static forces (even in the absence of any additional (delay) forces). For this reason, the application of the PT approach often requires additional approximations. In the present study we use two different approaches, which we will specify in Sec.\,\ref{SEC:linear-delay-force}.
\par
Yet another approach involves a small delay expansion on the level of the LE \cite{Guillouzic1999,Guillouzic2000, Longtin2010}. More specifically, the authors of Ref.\,\cite{Guillouzic1999,Guillouzic2000} suggest a Taylor expansion up to linear order of the total deterministic force (and of the noise intensity when delayed noise is considered) in powers of $\tau$. {The Taylor expansion is somewhat problematic, since it involves the derivative of the position $\chi$, which is a stochastic variable and hence not continuously differentiable. Performing this Taylor expansion \textit{ad hoc} hence implies a certain inaccuracy, which may become especially apparent in the range of large noise intensities (large $D_0$)}, see \cite{Frank2005} for a discussion.
\par
The novel approach introduced in the present paper works on the basis of the FPE, like the perturbation theory. As we will demonstrate, our ansatz provides an improved approximation of the one-time probability density for the considered nonlinear static forces.
\subsection{Force linearization closure\label{SEC:FLC} }
In the following, we focus on non-equilibrium steady states (NESS). Indeed, many Langevin systems described by Eq.\,(\ref{EQ:LE}) automatically reach a NESS, defined by ${\partial_t} \rho_{\mathrm{1,ss}}(x,t)\!=\!0$. Steady states generally allow for more analytical treatment than transients.
Throughout this work we denote steady state quantities with the subscript ``ss". Since the steady state conditional probability only depends on the time difference, and not on the instances of time themselves, we will further use the shortened notation: $\rho_{\mathrm{c,ss}}(x_{\tau}| x;\tau)=\rho_{\mathrm{c,ss}}(x_{\tau},t\!-\!\tau | x, t)$.
\par
The basis of our approach is that, within a NESS, the FPE hierarchy can be solved \textit{exactly} when all deterministic forces are linear in $x$ and $x_\tau$ \cite{Kuechler1992, Frank2001}. A further, yet less restrictive requirement is that the system obeys natural boundary conditions, i.e., $\lim_{x \to \pm \infty} \rho(x,t)\!=\!0$ and $\Omega\!=\!\mathbb{R}$. Given this background, the main idea of our approach is to achieve a closed approximate FPE for $\rho_1$ by {linearizing the deterministic forces} in all members of the infinite FPE hierarchy apart from the first one. Due to the involved linearization procedure, which we outline in detail below, we call our approach ``force linearization closure'' (FLC).
%
%
%
\subsubsection{Linearization of the deterministic forces\label{SEC:linearization-rule}}
We start by considering the (time-dependent) energy landscape resulting from the total force\,(\ref{EQ:F}). 
As a simple estimate for the total steady state energy landscape, we assume at this step the system to be at rest, i.e., $\chi(t-\tau)\!\equiv\!\chi(t)$, such that the total (static) potential is given by 
\begin{equation}\label{EQ:V_STAT}
V_\mathrm{STAT}(x) = -\int F(x,x_\tau\!=\!x) \mathrm{d}x.
\end{equation}
In general, $V_\mathrm{STAT}$ may have multiple (local) minima (defined by $F\!=\!0$ and $F'\!>\!0$). 
We number each minimum $x^i_{0}$ with an integer $i\in \! \mathcal{I}$. The index set $\mathcal{I}$ hence contains one element for each local minimum.
To formulate the FLC, we split the spatial domain $\Omega$ into non-overlapping intervals $\Omega^i$, such that both boundaries of $\Omega$ and each local maximum of $V_\mathrm{STAT}$ represents a bound of an interval, and every $\Omega^i$ contains exactly one minimum $x^i_{0}$.
\par
By this procedure, we obtain a splitting of $\Omega$ into subdomains $\Omega^i$, whose union is again the entire spatial domain: $\Omega\!=\! \bigcap _{{i \in \mathcal{I}}} \Omega^i$. For static potentials with a single minimum $x^j_{0}$ , this means $\Omega^j\!=\!\Omega$. For each $i$, we then perform a Taylor expansion of the entire deterministic force $F\!=\!F_\mathrm{s}+F_\mathrm{d}$ in both spatial variables $x$ and $x_\tau$. More specifically, we expand around the deviations with the enclosed minimum, i.e., $\Delta x^i\! =\! x-x^i_{0}$ and $\Delta x_{\tau} ^i\! =\! x_{\tau}-x^i_{0}$. Neglecting all terms of quadratic orders $\mathcal{O}(\Delta {x^{i2}})$, $\mathcal{O}(\Delta x^{i} \Delta x^{i}_\tau)$, $\mathcal{O}(\Delta{x_\tau^{i2}})$, 
or higher, we obtain
\begin{equation}\label{EQ:Force-linear}
F^{\mathrm{lin},i}(x,x_{\tau})\! =\!-{{\alpha}^i\Delta x^i} -{\beta}^i \Delta x^i_\tau,
\end{equation} 
where $-\alpha^i$ and $-\beta^i$ are the first-order derivatives of $F$ with respect to $x$ and $x_\tau$, respectively, evaluated at the minimum $x^i_{0}$. [Note that since we expand around the minima of $V_\mathrm{STAT}$ (\ref{EQ:V_STAT}), we always have $F(x^i_{0},x^i_{0})\!=\!0$, such that the constant terms vanish.]
\par
This linearization procedure yields an approximation of the steady state energy landscape composed of a sequence of quadratic polynomials with time-dependent centers, each subdomain $\Omega^i$ reaching from one local maximum to the following one (or to a bound of the entire spatial domain). 
%
%
\subsubsection{Analytical solution for linearized forces}
Now, we will turn back to the full system which involves thermal noise. Without further reasoning, one would expect that the linearization of the deterministic forces renders a good approximation of the stochastic dynamics, whenever the probability density close to the minima of the (approximate) total potential in the steady state [Eq.\,(\ref{EQ:V_STAT})] is large. This is, for instance, the case, when the potential barriers are high compared to thermal fluctuations.
\par
The following steps are performed separately for each subdomain $\Omega^i$. We first apply the linearization $F\approx F^{\mathrm{lin},i}$ [see Eq.\,(\ref{EQ:Force-linear})] to the force terms in all members of the infinite FPE hierarchy starting from the second, i.e., the equation for the two-time steady state density $\rho_{2, \mathrm{ss}}(x,t;x_\tau,t-\tau)$ (see Appendix \ref{APP:FPE_rho2} for the general from of this FPE). 
 Assuming natural boundary conditions at every subdomain bound, the FPEs for all $\rho_{\mathrm{n, ss}}$ can be solved by multivariate Gaussian distributions \cite{Kuechler1992}. The second member of the FPE hierarchy in the linear case and the corresponding analytical solutions $\rho^{\mathrm{lin}}_{\mathrm{2,ss}}$ and $\rho^{\mathrm{lin}}_{\mathrm{3,ss}}$ are given in the Appendices \ref{APP:FPE_rho2_linear} and \ref{APP:C}. For each subdomain, we thus have access to the function $\rho_{\mathrm{2,ss}}\equiv \rho^{\mathrm{lin},i}_{\mathrm{2,ss}}$ (the explicit formula for $\rho^{\mathrm{lin},i}_{\mathrm{2,ss}}$ can also be found in \cite{Frank2003}). The corresponding conditional probability density reads
\begin{align}\label{EQ:rho-c-ss_linear}
\rho^{\mathrm{lin},i}_{\mathrm{c,ss}}(x_{\tau}| x;\tau)=\sqrt{\frac{2 }{ \pi } K^i {d_1^i}^2 \left(1- {d_2^i}^2\right)}\exp\!\left[\frac{{\Delta x^i}^2}{2  K^i}\right]\times \nonumber \\ 
\exp\!\left[d^i_1\left( 2\Delta x^i\, \Delta x_{\tau}^id^i_2-{\Delta x^i}^2-{\Delta x_{\tau}^i}^2\right)\right]
\end{align}
with the coefficients
\begin{subequations}
\begin{align}
\omega^i \equiv &  \omega ({\alpha^i,\beta^i})\,\,\,\,\,\,=& \sqrt{(\alpha^{i})^2-(\beta^{i})^2}/ \gamma ~~~\in \mathbb{C},\\
K^i \equiv & \, K(\alpha^{i},\beta^{i},\tau)\, =&  D_0\frac{\gamma+(\beta^{i}/ \omega^{i}) \sinh(\tau \omega^{i} )}{\alpha^{i}+\beta^{i} \cosh(\tau \omega^{i})}, \label{EQ:K-tau}\\
d_1^i \equiv & \, d_1({\alpha^{i},\beta^{i},\tau})\, =&\frac{(\beta^{i})^2  K^i/2}{(\beta^{i} K^i)^2 - (D_0\gamma-\alpha^{i} K^{i})^2}, \label{EQ:d1} \\
d_2^i \equiv & \, d_2({\alpha^{i},\beta^{i},\tau})\,=&({D_0\gamma-\alpha^{i} K^{i}})/({\beta^{i} K^{i}}).\label{EQ:d2} 
\end{align}
\end{subequations}
We recall $\Delta x^i \!=\! x-x^i_{0}$, where $x^i_{0}$ is the enclosed minimum of $V_\mathrm{STAT}$ [Eq.\,(\ref{EQ:V_STAT})]. Note that $\omega^i$ becomes imaginary if $|\alpha^{i}|\!<\!\beta^{i}$, such that the hyperbolic functions in $K(\alpha^{i},\beta^{i},\tau)$ [see Eq.\,(\ref{EQ:K-tau})] convert to trigonometric ones. In this case, there exist critical $\tau$ values $\tau_c \omega=\arccos (-\alpha^{i}/\beta^{i})+2\pi \kappa, ~\forall \kappa \in \mathbb{Z} $, for which $K^i$ diverges, and no stable NESS exists. There is also no stable state, when $-\beta^{i}\leq \alpha^{i}$, or $\tau \alpha^{i}\!\leq\! -1$. Only if $-\alpha^{i}\!<\!\beta^{i}\!<\!\alpha^{i}$, a steady state is approached for all $\tau$. A derivation and discussion of the steady state conditions is presented in \cite{Kuechler1992}.
\par
With Eq.\,(\ref{EQ:rho-c-ss_linear}), the first member of the approximate FPE hierarchy (where the force linearization is applied to the second and all higher members) accordingly reads
\begin{subequations}\label{EQ:dFPE-DL}
\begin{align}\label{EQ:dFPE-DL1}
\displaystyle \gamma D_0 {\partial_{xx}} \rho^{\mathrm{FLC}}_{\mathrm{1,ss}}(x)  =   {\partial_x}[\widehat{F}^{\mathrm{FLC}}(x)\rho^{\mathrm{FLC}}_{\mathrm{1,ss}}(x) ],
\end{align}
where
\begin{align}
\displaystyle \widehat{F}^{\mathrm{FLC}}(x)=\int_{\Omega}\!\!  F(x,x_\tau\!)\rho_{\mathrm{c,ss}}^{\mathrm{lin},i}(x_{\tau}| x;\tau) \mathrm{d}x_{\tau}, ~\forall x \in \Omega^i .\label{EQ:dFPE-DL2} 
\end{align}
\end{subequations}
In Eq.\,(\ref{EQ:dFPE-DL1}), $\rho^{\mathrm{FLC}}_{\mathrm{1,ss}}$ is the steady-state one-time probability density obtained within the FLC approach. Notice that the {deterministic force $F$} appearing in the integral in Eqs.\,(\ref{EQ:dFPE-DL2}) is \textit{not} linearized.
\par
Equations\,(\ref{EQ:rho-c-ss_linear}-\ref{EQ:dFPE-DL2}) [together with the linearization rule (\ref{EQ:Force-linear})] form a closed set of equations. The here presented closure of the FPE hierarchy is the cornerstone of our approach. Since we have a general expression for $\rho_{\mathrm{c,ss}}^{\mathrm{lin},i}$ [Eq.\,(\ref{EQ:rho-c-ss_linear})] with which $\widehat{F}^{\mathrm{FLC}}$ can be calculated separately, the FLC formally converts the delayed FPE for $\rho_\mathrm{1,\mathrm{ss}}$ itself into a closed, {quasi-Markovian} one [Eq.\,(\ref{EQ:dFPE-DL1})]. The fact that the delay is still present in our approximate FPE, becomes apparent by considering the special case of a linear force $F$. Then, Eqs.\,(\ref{EQ:dFPE-DL}) coincide with the corresponding exact \textit{delayed} FPE. Furthermore, one can easily see from Eqs.\,(\ref{EQ:dFPE-DL}) that the usual FPE is recovered when the delay time vanishes, or when $F$ becomes independent of $x_\tau$: in both cases $\widehat{F}^{\mathrm{FLC}}\!=\!F$.
\subsubsection{Vanishing steady state probability current}
The key equations of the FLC [Eqs.\,(\ref{EQ:rho-c-ss_linear}-\ref{EQ:dFPE-DL2})] are valid for steady states of, in principle, {arbitrary systems which can be meaningfully linearized according to the procedure }described 
in Sec.\,\ref{SEC:linearization-rule}.
A particularly simple situation arises, if additionally the steady-state probability current given by $ J\! =\! [\widehat{F}^{\mathrm{FLC}}(x)- \gamma D_0 {\partial_x}]\rho^{\mathrm{FLC}}_{\mathrm{1,ss}}(x)$ vanishes.
If $J=0$, the formal solution of Eq.\,(\ref{EQ:dFPE-DL1}) takes the simple Boltzmann-like form
 \begin{equation}\label{EQ:formal-sol-vanishingJss}
 \rho^{\mathrm{FLC}}_{\mathrm{1,ss}}(x)  ={Z}^{-1} \exp\left[-{V_{\mathrm{eff}}^{\mathrm{FLC}}(x)}/({  k_\mathrm{B} \mathcal{T}  }) \right]
 \end{equation} 
with the effective static potential
\begin{equation}\label{EQ:Veff}
V_{\mathrm{eff}}^{\mathrm{FLC}}(x)=-\int_{\hat{x}}^x \mathrm{d}x'\widehat{F}^{\mathrm{FLC}}(x'),
\end{equation} 
where $\widehat{F}^{\mathrm{FLC}}$ is given in Eq.\,(\ref{EQ:dFPE-DL2}), and $\hat{x} \in \Omega$ is arbitrary but fixed.
Here and in the following, we denote the normalization constant by $Z$.
\par 
We stress that the assumption $J\!=\!0$ simplifies the analysis, but it is not a necessary condition. Also for non-zero currents, Eq.\,(\ref{EQ:dFPE-DL1}) can be treated using standard techniques for (Markovian) Fokker-Planck equations \cite{Risken1984, Gardiner2002}.
\section{Applications\label{SEC:application}}
We now apply the FLC to two generic examples, involving a multistable and a bistable static potential combined with a linear delay force. In the subsequent section~\ref{SEC:linear-delay-force}, we first define this force and provide some results which apply to both model systems. In Secs.\,\ref{SEC:washboard} and \ref{SEC:doublewell} we then present results from the FLC and compare its performance to the two main approaches known from the literature, i.e., the perturbation theory (PT) and the small delay expansion. As a test of the different approximations, we provide numerical results from Brownian dynamics (BD) simulations. Details about the numerical methods are given in the Appendix \ref{APP:BD}. 
\subsection{Linear delay force, general results}\label{SEC:linear-delay-force}
We consider a linear delay force, with amplitude $k\ge0$,
\begin{equation}\label{EQ:F_delay}
F_\mathrm{d}(x,x_\tau)\! =\! -k(x-x_\tau), 
\end{equation}
which vanishes for $k\!\rightarrow\!0$ or $\tau\!\rightarrow\!0$. The particular force in Eq.\,(\ref{EQ:F_delay}) can be associated with the delayed confining potential $V_\mathrm{d}(x,x_\tau)=(k/2)[x-x_\tau]^2$. Such quadratic feedback potentials are commonly used to model optical traps \cite{Kotar2010, Kim2007, Woerdemann2012}, which are implemented in many experimental setups to control colloidal particles \cite{Kotar2010, Kim2007, Qian2013,*Balijepalli2012, Balijepalli2012}. From a more general perspective, the delay force in Eq.\,(\ref{EQ:F_delay}) is of Pyragas type \cite{Pyragas1992}, and has been extensively studied in the context of chaos control \cite{Schoell2008}. It is important to note that the FLC generally also applies to systems with nonlinear delay forces.
\par
For both exemplary systems, we use natural boundary conditions. Therewith, the steady state probability current vanishes irrespective of {the particular form of} the static potential $V_\mathrm{s}$, and solution (\ref{EQ:formal-sol-vanishingJss}) readily applies. We can give a general expression for $\rho_\mathrm{1,ss}^{\mathrm{FLC}}$ as follows. The delay force is already linear, and yields $\beta^i\!=\!-k, \forall i \in \mathcal{I}$, where $\beta^i$ and $k$ are the coefficients appearing in Eqs.\,(\ref{EQ:Force-linear}) and (\ref{EQ:F_delay}), respectively. After performing several Gaussian integrals, Eq.\,(\ref{EQ:dFPE-DL2}) yields
\begin{align}
\widehat{F}^{\mathrm{FLC}}(x)\!=\! -\partial_x V_{\mathrm{s}}(x) -k \Delta x^i+kx {d^i_2}\sqrt{2 K^i d^i _1 }\, \times \nonumber \\
\sqrt{1-{d^i_2}^2} \,\exp \left\{ \left[ d^i_1 {d^i_2}^2-d^i _1+{1}/({2K^i } )\right] { \Delta x^i} ^2 \right\}.
\end{align}
$\widehat{F}^{\mathrm{FLC}}$ can be further simplified by using the identity 
\begin{equation}\label{EQ:d1-d2-identity}
d^i_1{d^i_2}^2-d^i_1+{1}/({2K^i}) =0,
\end{equation}
where the quantities $d^i_1$ and ${d^i_2}$ are given in Eqs.\,(\ref{EQ:d1}) and (\ref{EQ:d2}). This
yields the piecewise defined effective potential
\begin{align}\label{EQ:Veff-FLC}
& V_{\mathrm{eff}}^{\mathrm{FLC}}(x)=V_{\mathrm{s}}(x) + \textcolor{black}{ {k}/{2}\!\left( 1- |d^i_2| \right) } { \Delta x^i} ^2,
\end{align}
for $x \in \Omega^i$. Inserting Eq.\,(\ref{EQ:Veff-FLC}) into  Eq.\,(\ref{EQ:formal-sol-vanishingJss}) one obtains $\rho^{\mathrm{FLC}}_{\mathrm{1,ss}}$. Furthermore, since $F_\mathrm{d}(x,x)\!=\!0\,\forall x\in \Omega$, our estimate for the steady state energy landscape coincides with the static potential $V_\mathrm{STAT}\!=\!V_\mathrm{s}$, and hence $x^i_{0}$ and the bounds of $\Omega^i$ are readily determined by the extrema of $V_\mathrm{s}$.
Thus, for a given $V_{\mathrm{s}}$, one only needs to calculate $\alpha^i$ by linearizing $F_\mathrm{s}$, and therewith the coefficient $d^i_2$ [Eq.\,(\ref{EQ:d2})].
\par
Before specifying $V_\mathrm{s}$, we review some results from the two main earlier approaches, for the case of the linear delay force [Eq.\,(\ref{EQ:F_delay})] (and natural boundary conditions). Just like the FLC [Eq.\,(\ref{EQ:formal-sol-vanishingJss})], the small delay expansion gives rise to a Boltzmann-distributed one-time probability density
\begin{equation}\label{EQ:BoltzmannDis}
 \rho^{\mathrm{approx}}_{\mathrm{1,ss}}(x)  ={Z}^{-1} \exp\left[-{V_{\mathrm{eff}}^{\mathrm{approx}}(x)}/({  k_\mathrm{B} \mathcal{T}  }) \right] .
\end{equation}
Here, the effective potential $V_{\mathrm{eff}}^{\mathrm{approx}}\equiv V^{\mathrm{small}\tau}_\mathrm{eff}$ reads \cite{Frank2005}
\begin{equation}\label{EQ:Veff-smallTau}
V^{\mathrm{small}\tau}_\mathrm{eff}(x)= V_\mathrm{s}(x)/(1-k\tau).
\end{equation}
The application of the perturbation-theoretical approach, on the other hand, is not straightforward. The closed approximate FPE from the PT involves the conditional probability of the corresponding unperturbed ($F_\mathrm{d}\!\equiv\! 0$) system, $\rho^{F_d\equiv0}_{\mathrm{c,ss}}(x_\tau|x;\tau)$. Since for most static potentials (including the ones we will consider in Secs.\,\ref{SEC:washboard} and \ref{SEC:doublewell}) no analytical expression for this quantity is available, further approximations become inevitable. As suggested in Ref.\,\cite{Frank2005}, one can for this purpose utilize the ``short time propagator'' \cite{Risken1984}
\begin{align}\label{EQ:Short-time-prop}
\rho_{\mathrm{c}}(x_{\tau}&| x;\tau)= \nonumber \\
&\frac{1 }{\sqrt{4 \pi D_0 \tau} }\exp\!\left[-\frac{{\{x_\tau - x - F_\mathrm{s}(x)\tau/\gamma \} }^2}{4 D_0 \tau}\right],
\end{align}
which is derived and discussed in Ref.\,\cite{Risken1984} on p.\,73 and prior. The usage of the short time propagator implies a first-order approximation in $\tau$ \cite{Risken1984, Frank2005}. The resulting one-time probability density is also the Boltzmann distribution Eq.\,(\ref{EQ:BoltzmannDis}) with $V_{\mathrm{eff}}^{\mathrm{approx}}\equiv V^{\mathrm{PT}}_\mathrm{eff}$ given by
\begin{equation}\label{EQ:Veff_PT}
V^{\mathrm{PT}}_\mathrm{eff}(x)=(1+k\tau) V_\mathrm{s}(x).
\end{equation}
Interestingly, when the short time propagator is used {for $\rho^{F_d\equiv0}_{\mathrm{c,ss}}$}, the PT approach obviously yields qualitatively very similar results as the small $\tau$ expansion. In particular, both approaches render effective potentials that are proportional to the static potential [Eqs.\,(\ref{EQ:Veff-smallTau}) and (\ref{EQ:Veff_PT})]. This means, also the approximate densities have the same functional form as in the case $F_\mathrm{d}\!\equiv\! 0$. In fact, the only remaining effect of the delay force, according to these approximations, is a constant ($x$-independent) factor within the exponent of $\rho^\mathrm{approx}_\mathrm{1,ss}$. Moreover, the small $\tau$ expansion and the PT with short time propagator even become equivalent for small $\tau k$, since then these factors coincide $1/(1-k\tau)=(1+k\tau) +\mathcal{O}([\tau k]^2)$.
\par
 Alternatively, we propose to approximate $\rho^{F_d\equiv0}_{\mathrm{c,ss}}$ by the conditional probability from the corresponding unperturbed ($F_d\equiv0$), linearized system, i.e., the corresponding Ornstein–Uhlenbeck process, given in Ref.\,\cite{Risken1984} (p.\,100). The resulting effective potential from the PT approach thus reads
 \begin{align}\label{EQ:Veff_PT-OU}
V^{\mathrm{PT}}_{\mathrm{eff}}(x)=& V_\mathrm{s}(x) + (k/2)\times \nonumber\\
 &\!\left\{ 1- \exp{[-(\alpha^i\!-\!k)\tau /\gamma]} \right\} {\Delta x^{i}}^2, \forall x\in \Omega^i.
 \end{align}
As opposed to the result obtained with the short-time propagator [Eq.\,(\ref{EQ:Veff_PT})], the effective potential in Eq.\,(\ref{EQ:Veff_PT-OU}) has a different functional form than $V_\mathrm{s}(x)$. More specifically, the delay force now effectively adds to the static potential a fixed quadratic potential around the minima of $V_\mathrm{s}$. Since the linear delay force [Eq.\,(\ref{EQ:F_delay})] is indeed expected to trap the particle in a quadratic confining potential (with history-dependent center), this effective potential appears to be more realistic than Eq.\,(\ref{EQ:Veff_PT}). However, the history-dependency of the {position of the} ``trap'' imposed by $F_\mathrm{d}$ is, of course, also not captured by this approximations.
 \par
 As a first consistency check, one readily sees from Eqs.\,(\ref{EQ:Veff-FLC}-\ref{EQ:Veff-smallTau}),(\ref{EQ:Veff_PT}),(\ref{EQ:Veff_PT-OU}) that all four approximations give rise to Boltzmann-distributed density profiles, which, in the limits of vanishing delay force ($\tau \rightarrow 0$ or $k \rightarrow 0$), all coincide with the exact result for the non-delayed (Markovian) FPE, i.e., $\rho_\mathrm{1,ss}\propto \exp [-{V_\mathrm{s}(x)/k_\mathrm{B}\mathcal{T}}]$. However, they yield different effective potentials for finite delay forces, which we will compare in the following.
 \par
Finally, we note that considering the force (\ref{EQ:F_delay}) rather than a force with two different amplitudes $F_\mathrm{d}(x,x_\tau)\! =\! -k' x- k x_\tau,$ does not imply a loss of generality concerning the results in this section. A minor adjustment required if $k'\! \neq \! k$, is that $x^i_{0}$ are then the minima of $V_\mathrm{STAT}\neq V_\mathrm{s}$. 
\subsection{Periodic static potential}\label{SEC:washboard}
We start by considering the periodic ``washboard'' potential
\begin{equation}\label{EQ:Vs-washboard}
V_{\mathrm{s}}(x)\!=\!-(\Delta \! V_\mathrm{ }/2) \cos (x/ \sigma)
\end{equation}
with barriers of height $\Delta \! V_\mathrm{ }$ and minima at $x^i_{0}\!=\!2\pi\sigma i$ and $i\!\in \!\mathcal{I}\!=\!\mathbb{Z}$. Particles in sinusoidal potentials represent a well-studied paradigm to model transport on rough surfaces \cite{Juniper2016}.
Linearizing the static potential yields  $\alpha^i \!=\! \Delta \! V_\mathrm{ }/(2\sigma^2)+k$ (and $\beta^i\! =\! -k$) for all $i \in\mathcal{I}$. On all subdomains 
$\Omega^i\!=\![(2i-1) \pi\sigma , (2i+1)\pi\sigma]$, the density is hence given by the Boltzmann distribution (\ref{EQ:formal-sol-vanishingJss}) with effective potential (\ref{EQ:Veff-FLC}), 
where $V_\mathrm{s}$ is given in Eq.\,(\ref{EQ:Vs-washboard}) and
\begin{equation}\label{EQ:rho_washboard_FLC}
d^i_2=\frac{\gamma\omega\cosh(\tau \omega )-[{\Delta \! V_\mathrm{ }}/({2\sigma^2})+k]\sinh(\tau \omega )}{\gamma\omega-k \sinh(\tau \omega )}
\end{equation}
with $\gamma\omega\sigma=\sqrt{({\Delta \! V_\mathrm{ }}/{2\sigma})^2+k \Delta \! V_\mathrm{ }}$ and $\Delta x^i = x-x^i_{0}$.
\par
\begin{figure}
\includegraphics[width=\linewidth]{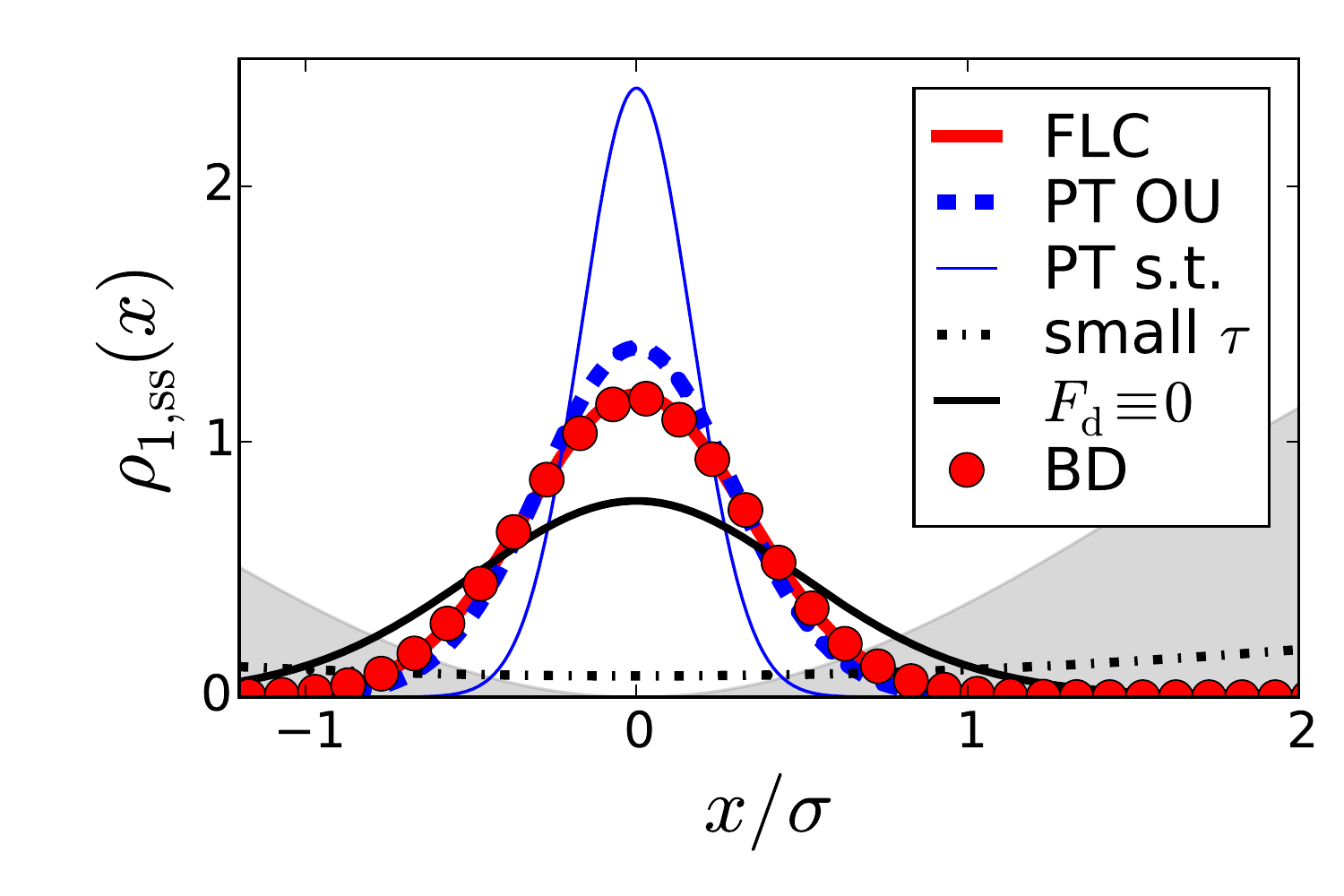}
\caption{(Color online)
One peak of the periodic steady state probability density $\rho_\mathrm{1,ss}$ in the ``washboard'' with delay force. Red symbols: numerical data (BD), 
thick red line: force linearization closure (FLC), blue dashed line: perturbation theory with Ornstein–Uhlenbeck approximation (PT OU), blue thin line: PT with short time propagator (PT s.t.), dashed-dotted line: small delay expansion (small $\tau$), and solid black line: density without delay force $\rho_\mathrm{1,ss}(x)\propto \exp\{-V_\mathrm{s}(x)/(k_\mathrm{B}\mathcal{T})\}$ ($F_\mathrm{d}\!\equiv \!0$). The delay force amplitude and delay time are set to $k\!=\!8\,k_{\mathrm{B}}{\cal T}/\sigma^2 $ and $\tau\!=\!\tau_{\mathrm{B}}$, respectively. Here and in the following plots the barrier height is set to $\Delta \! V_\mathrm{ }\!=\!8\,k_{\mathrm{B}}{\cal T}$.
}
\label{FIG:rho}
\end{figure}
\begin{figure} 
\includegraphics[width=\linewidth]{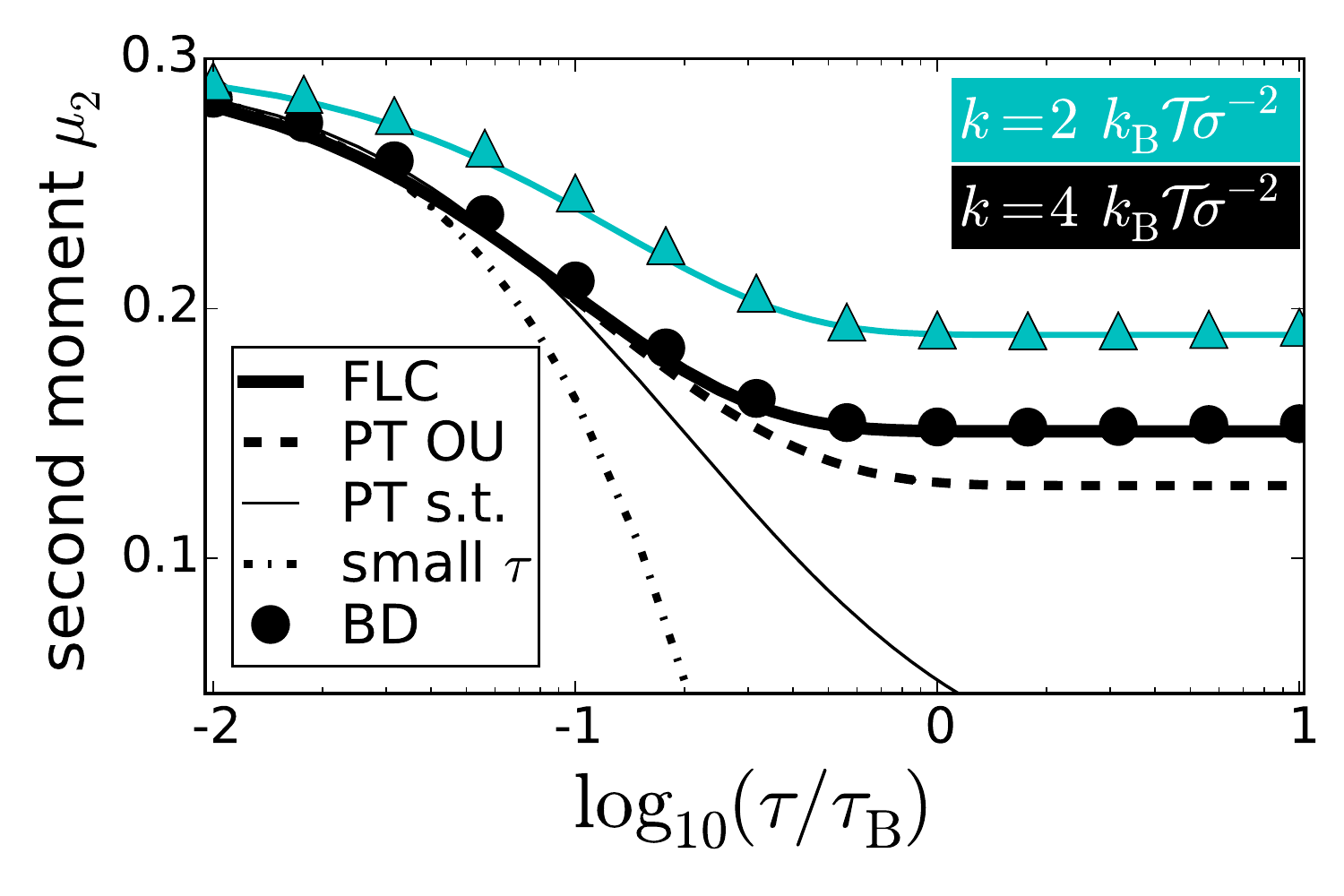}
\caption{(Color online) Second moment $\mu_2$ of the steady state probability density within one {well} of the ``washboard'' potential, e.g., of $\rho_\mathrm{1,ss}(-\pi\sigma\! \le\! x\! \le \! \pi\sigma)$. Symbols show results from Brownian dynamics (BD), and thick solid lines from force linearization closure (FLC). Cyan color is $k\!=\!2\,k_{\mathrm{B}}{\cal T}/\sigma^2$. Black is $k\!=\!4\, k_{\mathrm{B}}{\cal T}/\sigma^2$, with thin solid line: perturbation theory with short time propagator (PT s.t.), dashed line: perturbation theory with Ornstein–Uhlenbeck $\rho^{F_\mathrm{d}\equiv0}_\mathrm{c,ss}$ (PT OU), and dot-dashed line: small $\tau$ Taylor expansion (small $\tau$).
}\label{FIG:washboard_BD_PT_VAR}
\end{figure}

Figure~\ref{FIG:rho} shows the density profile in one potential well generated with BD simulations of the delayed LE (\ref{EQ:LE}), on the one hand, and the corresponding FLC approximation (\ref{EQ:rho_washboard_FLC}), on the other hand, for an exemplary parameter choice. Here and throughout the entire paper, we set the barrier heights to $\Delta \! V_\mathrm{ }\!=\!{8}\, k_{\mathrm{B}}{\cal T}$. One clearly sees that the FLC generates a very good approximation of the probability density distribution. Similar convincing results are found for all other considered values of $\tau$ and $k$ in the tested range $\tau/\tau_\mathrm{B}\in [10^{-2},10]$ and $k/(\sigma^{-2}k_\mathrm{B}\mathcal{T}) \in [0,32]$. To compare approximation and simulation results more systematically, we calculate the moments $\mu_n$ of the distributions within one potential well (for $x$ values between two maxima of $V_\mathrm{s}$), e.g., of $\rho_\mathrm{1,ss}(-\pi\sigma\le x \le \pi\sigma)$. Due to the spatial inversion symmetry of the total potential{ around the enclosed minimum}, all odd moments vanish. We hence consider the second central moment $\mu_2\!=\! \langle (\chi - \mu_1)^2 \rangle_\mathrm{ss}$ (with $\mu_1\!=\!\langle \chi  \rangle_\mathrm{ss}$){ of one peak of the distribution}. Figure~\ref{FIG:washboard_BD_PT_VAR} shows $\mu_2$ vs. the delay time $\tau$ for two exemplary values of delay force amplitude. We find that in the presence of the delay force (\ref{EQ:F_delay}), the variance of the distribution is reduced. This is not surprising, since the delay force arises from a quadratic potential which ``traps'' the particle and hence enhances the confining effect of $V_\mathrm{s}$. Within the considered parameter regime, the variance is seen to decrease with increasing $\tau$, until a saturation value is approached at $\tau^\mathrm{W}_\mathrm{sat}\! \approx \! 0.5\,\tau_ \mathrm{B}$ (due to the strongly increasing simulation times, we did not simulate much larger $\tau$).
\par
On the level of the probability densities and the FPE, the saturation of $\mu_2$ at large $\tau$ indicates that in the considered range of delay times, the delay-averaged drift $\widehat{F}$, and hence $\rho_\mathrm{c,ss}(x_\tau|x;\tau)$, are essentially constant for $\tau\!>\!\tau^\mathrm{W}_\mathrm{sat}$. In other words, the conditional probability for the time difference $\tau$ remains essentially unchanged, when $\tau$ is further increased. This suggests some kind of relaxation mechanism within the valley, where the history of the stochastic process becomes ``washed out'' on the level of ensemble averaged quantities, like the probability densities. In this context it is interesting to note that, for the case without delay force ($k\!=\!0$), the relaxation time within a potential {well} is of the order $\tau^{\mathrm{W}}_{\mathrm{ir}}\!\approx\! {\gamma}/{V_\mathrm{s}''(x_\mathrm{min})}= 2\, \tau_\mathrm{B}k_\mathrm{B} \mathcal{T} /\Delta \! V_\mathrm{ }$ (see Ref.\,\cite{Gardiner2002}, p.\,348). In the present case, $\Delta \! V_\mathrm{ }\!=\!{8}\, k_{\mathrm{B}}{\cal T}$, that is $\tau^{\mathrm{W}}_{\mathrm{ir}}\!\approx\! 0.25\, \tau_\mathrm{B}$. The saturation of $\mu_2$ thus sets in \textit{after} the density relaxation within a potential well, consistent with our expectation.
\par 
A different situation occurs for much larger $\tau$ values than the ones considered in Fig.\,\ref{FIG:washboard_BD_PT_VAR}, in particular, when $\tau$ gets into the range of the mean escape times {(numerical results for the escape times are provided in Sec.\,\ref{SEC:Escape-times})}. Then we expect again a $\tau$-dependency of $\rho_\mathrm{c,ss}$ and therewith of $\mu_2$, since the transport between the potential valleys becomes important.
\par
We now compare the density from the FLC approach with corresponding data from the small delay (Taylor) expansion [Eq.\,(\ref{EQ:BoltzmannDis}) and (\ref{EQ:Veff-smallTau})] and from the PT approach. Within the latter, we either use the short time propagator [see Eq.\,(\ref{EQ:Veff_PT})], or, the corresponding Ornstein–Uhlenbeck approximation [see Eq.\,(\ref{EQ:Veff_PT-OU}) and above]. As visible in Fig.\,\ref{FIG:rho} and \ref{FIG:washboard_BD_PT_VAR}, the PT generally overestimates the height of the density peak. Approximating the conditional probability with the short time propagator yields worse results than using $\rho_\mathrm{c,ss}$ of the corresponding Ornstein–Uhlenbeck process. This observation matches our expectations, see Sec.\,\ref{SEC:linear-delay-force}. At least when the Ornstein–Uhlenbeck approximation is used, the PT reproduces the quantitative behavior of the function $\mu_2(\tau)$, see Fig.\,\ref{FIG:washboard_BD_PT_VAR}. On the contrary, the small delay expansion fails completely for larger delay times.
\par
 We conclude that the FLC generates the best approximation of the steady-state density distribution in the periodic potential, especially in the regime of large $\tau$ or {large} $k$. This is the regime where the ``perturbation'', i.e. $|F_\mathrm{d}|$, is not small compared to the force applied by the static potential, such that the basic assumption of the perturbation theory is not fulfilled. The small delay expansion, on the other hand, involves a truncated Taylor expansion in $\tau$, making it plausible that also this approach fails for large delay times.
\subsection{Bistable static potential}\label{SEC:doublewell}
As a second example, we consider the double-well potential
\begin{equation}\label{EQ:Vs_doublewell}
V_\mathrm{s}(x)\!=\! \Delta \! V_\mathrm{ } \,[  \left({x}/{\sigma}\right)^4 -2  \left({x}/{\sigma} \right)^2 ],
\end{equation}
with a barrier of height $\Delta \! V_\mathrm{ }$ (set to $\Delta \!V\!=\!8\, k_{\mathrm{B}}{\cal T}$) at $x/\sigma\!=\!0$ and two minima at $x^i_{0}=i\sigma$
with $i \in \mathcal{I}\!=\! \{-1,1\}$.
\par
A Brownian particle in a double-well potential is a generic model for a bistable noisy system \cite{Risken1984, Gardiner2002}, which in recent years has been considered also under the impact of a linear delay force \cite{Tsimring2001, Masoller2003, Du2015, Xiao2016, Piwonski2005}.
As opposed to the corresponding expression given in Eq.\,(\ref{EQ:F_delay}) (which corresponds to a moving ``optical tweezer''), these earlier theoretical studies consider a slightly different delay force involving solely the delayed particle position, i.e., $F_\mathrm{d}(x,x_\tau)= - kx_\tau$. Our approach can, however, as well be applied to such a delay force.
\par
The force linearization (according to the procedure described in Sec.\,\ref{SEC:linearization-rule}) yields
$\alpha^i \!=\! 8\,\Delta \! V_\mathrm{ }/\sigma^2+k$ for both $i$ (and $\beta^i\! =\! -k$). For $x/\sigma\in\![-\infty,0]$ and $x^{i=-1}_{0}\!=\!-\sigma$, or $x/\sigma \in \![0,\infty]$ and $x^{i=1}_{0}\!=\!\sigma$, the effective potential $V^\mathrm{FLC}_\mathrm{eff}$ is given by Eq.\,(\ref{EQ:Veff-FLC}) with
\begin{align}
 d^{i}_2=&\frac{\gamma\omega\cosh(\tau \omega )-[({8\Delta \! V_\mathrm{ }}/{\sigma^2})+k]\sinh(\tau \omega )}{\gamma\omega-k \sinh(\tau \omega )}
\end{align}
with $\gamma\omega\sigma=\sqrt{({8\Delta \! V_\mathrm{ }}/{\sigma})^2+k \Delta \! V_\mathrm{ }}$.
\par
\begin{figure} 
\includegraphics[width=\linewidth]{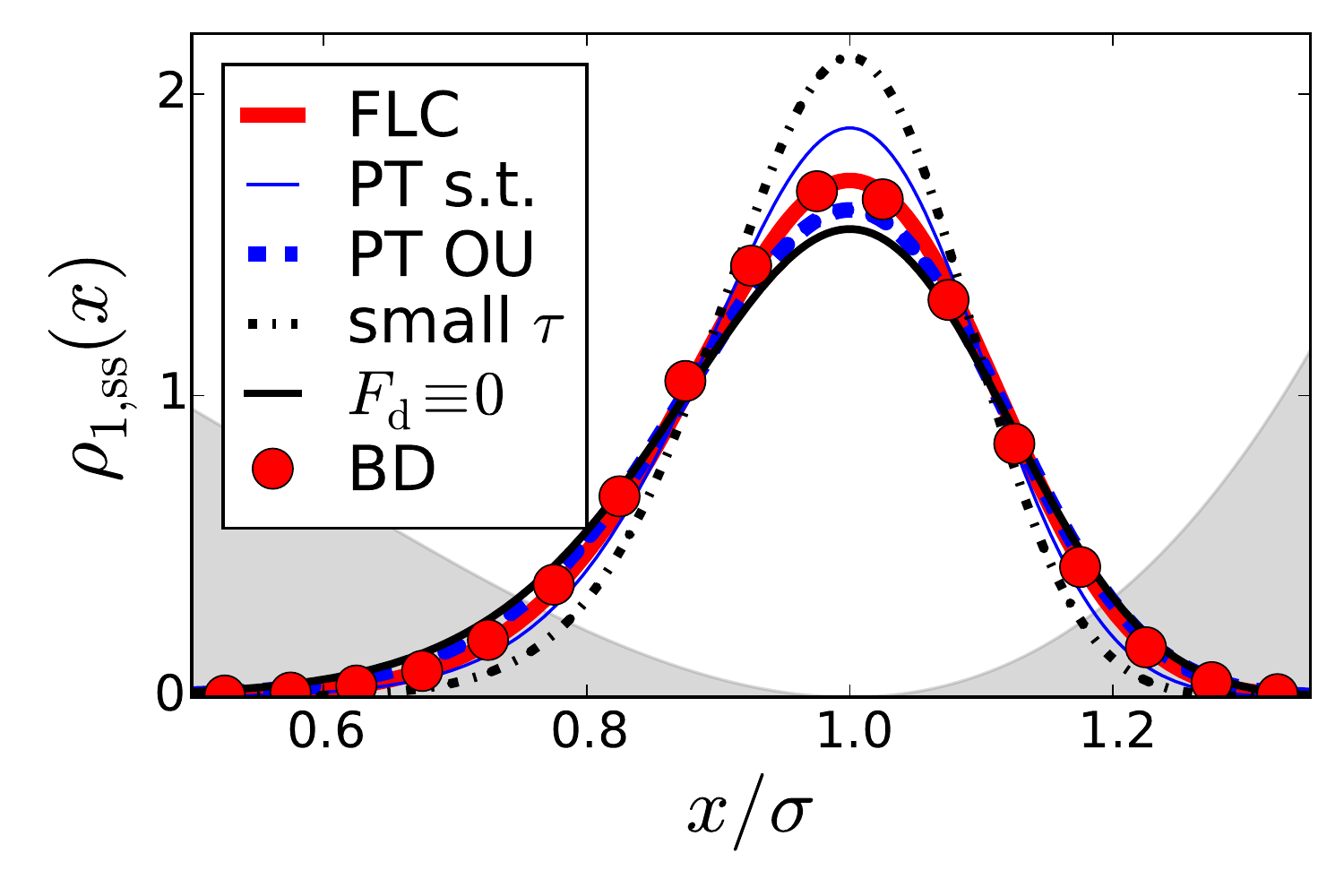} 
\caption{(Color online) One peak of the bimodal steady state probability density $\rho_\mathrm{1,ss}$ in the double-well potential. 
Color code as in Fig.\,\ref{FIG:rho}. The delay force amplitude and delay time are set to $k\!=\!18\,k_{\mathrm{B}}{\cal T}/\sigma^2 $ and $\tau\!=\!{0.1}\,\tau_{\mathrm{B}}$, respectively.
} 
\label{FIG:rho-DL}
\end{figure}
\begin{figure} 
\includegraphics[width=\linewidth]{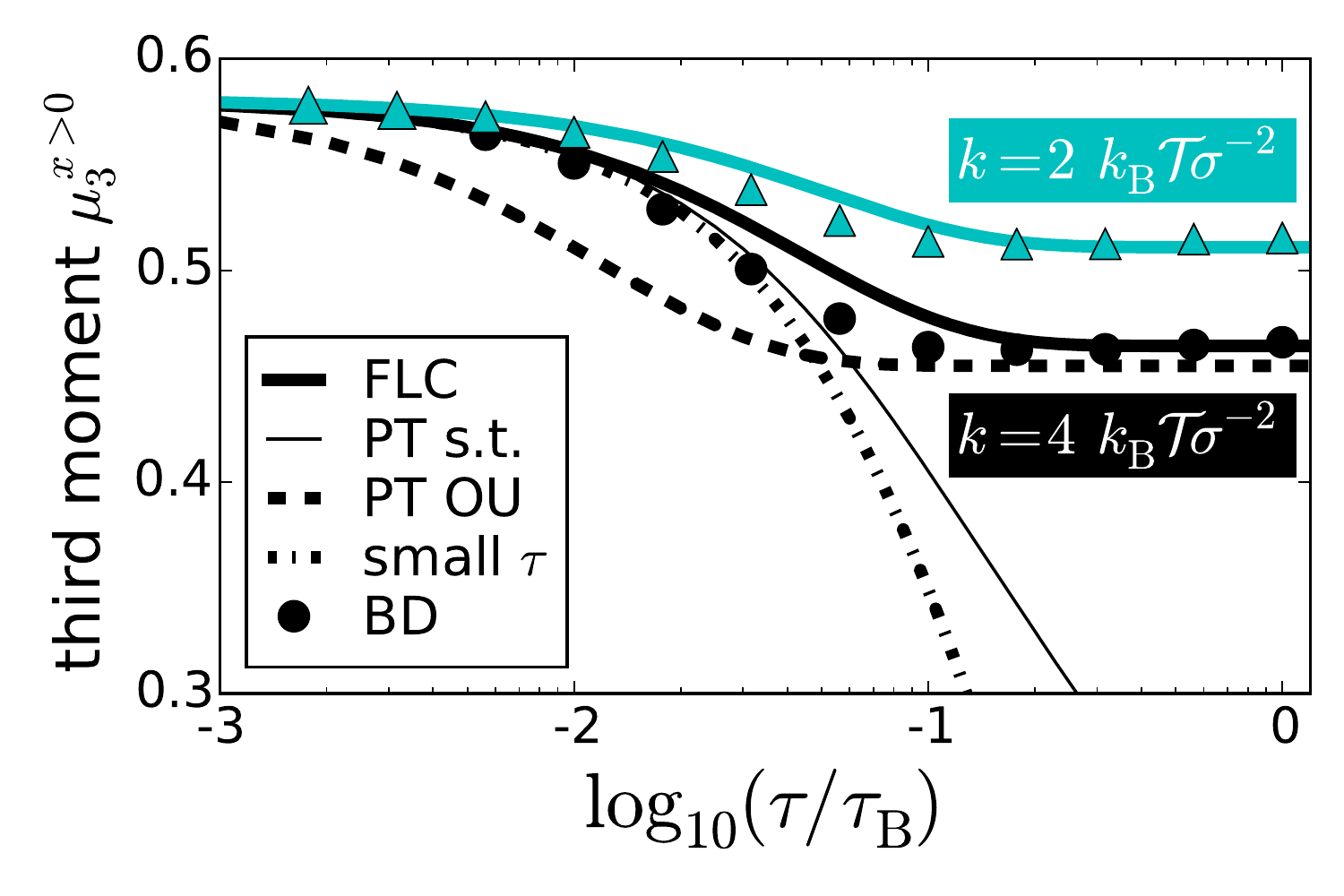}
\caption{(Color online) Third moment $\mu^{x>0}_3$ of the steady state probability density within the right well of the double-well potential, i.e., of $\rho_\mathrm{1,ss}(x\!>\!0)$. 
Color code and parameters as in Fig.\,\ref{FIG:washboard_BD_PT_VAR}. 
}
\label{FIG:mu3}
\end{figure}

Also for this model, the FLC clearly renders a very good approximation of the one-time probability density, as shown in Fig.\,\ref{FIG:rho-DL} for an exemplary parameter choice. Because the second moment is barely affected by the delay force, we here consider the third moment of one peak of the bimodal distribution, i.e., $\rho_\mathrm{1,ss}(x\!>\!0)$, as a function of the delay time. More specifically, we use the skew $\mu^{x>0}_3\!=\! \int_{x>0} [(x - \mu_1)/\sqrt{\mu_2}]^3 \rho_{1,\mathrm{ss}}(x) \mathrm{d}x $, i.e., the third central moment normalized with the standard deviation. The comparison with the numerical data in Fig.\,\ref{FIG:mu3} reveals that the FLC approach renders reasonably good predictions of $\mu^{x>0}_3$. We moreover find convincing results for all other tested parameters within the considered ranges $\tau/\tau_\mathrm{B}\in [10^{-2},5]$ and $k/(\sigma^{-2}k_\mathrm{B}\mathcal{T}) \in [0,18]$. Please note that this range spans from rather small delay forces that merely affect the shape of the energy landscape, to $k$ values so high that there is no second minimum of the total potential, when $\chi(t-\tau)$ rests in the first minimum. For all these qualitatively different cases, the agreement between FLC and numerical results is good. However, it is not as accurate as that for the second moment in the case of the periodic potential (see Fig.\,\ref{FIG:washboard_BD_PT_VAR}). One reason for that might be the fact that the approximate potential is, by construction, symmetric with respect to the {enclosed} minimum (within each subdomain). In this sense, the double-well potential, which is asymmetric around each minimum, is not as well approximated as the symmetric ``washboard'' potential. This manifests in a larger magnitude of the higher order Taylor terms that are neglected within the linearization procedure\,(\ref{EQ:Force-linear}). We also note that the third moment is \textit{per se} very sensitive against inaccuracies, since it involves cubic terms in the relative position. The deviations between the FLC and the exact result are hence expected to be larger.
\par
Regarding the impact of increasing $k$ and $\tau$, we see from Fig.\,\ref{FIG:mu3} that the skew $\mu^{x>0}_3$ becomes smaller, which means that the probability distribution becomes more symmetric. This is not surprising since the pure delay potential is quadratic (i.e., symmetric) in the system state variable. Moreover, similar to the second moment in the case of the ``washboard'' potential (see Fig.\,\ref{FIG:washboard_BD_PT_VAR}), $\mu^{x>0}_3$ approaches a saturation value (within the considered parameter range). The corresponding delay time is about one order of magnitude smaller than in the ``washboard'' case: $\tau^\mathrm{D}_\mathrm{sat}\!\approx \!0.1\tau_\mathrm{B}$ for $\Delta \! V_\mathrm{ }\!=\!8 \,k_\mathrm{B}\mathcal{T}$. 
This value of $\tau^\mathrm{D}_\mathrm{sat}$ is, as in the case of the ``washboard'', significantly larger than that for the relaxation time $\tau^{\mathrm{D}}_{\mathrm{ir}}$ within a well. For the bistable system at $k\!=\!0$ one finds \cite{Gardiner2002} (see p.\,348) $\tau^{\mathrm{D}}_{\mathrm{ir}}/\tau_\mathrm{B}\!\approx\! 1/8 \,(k_\mathrm{B} \mathcal{T} /\Delta \! V_\mathrm{ })=1/64 $ at $\Delta \! V_\mathrm{ }=8\,k_\mathrm{B} \mathcal{T}$. Thus, the saturation of $\mu_3$ can be explained by the same arguments as in the case of the ``washboard'' potential.
\par
Figure\,\ref{FIG:rho-DL} and \ref{FIG:mu3} also show the results according to the PT [Eqs.\,(\ref{EQ:Veff_PT}) and (\ref{EQ:Veff_PT-OU})], and from small delay expansion, see Eq.\,(\ref{EQ:Veff-smallTau}). Very similar to the case of the ``washboard'', only the PT with Ornstein–Uhlenbeck approximation is capable of reproducing the quantitative behavior (especially for large $\tau$), and the FLC clearly provides the best approximation of the one-time probability.
\subsection{Escape times \label{SEC:Escape-times}}
\begin{figure} 
\includegraphics[width=\linewidth]{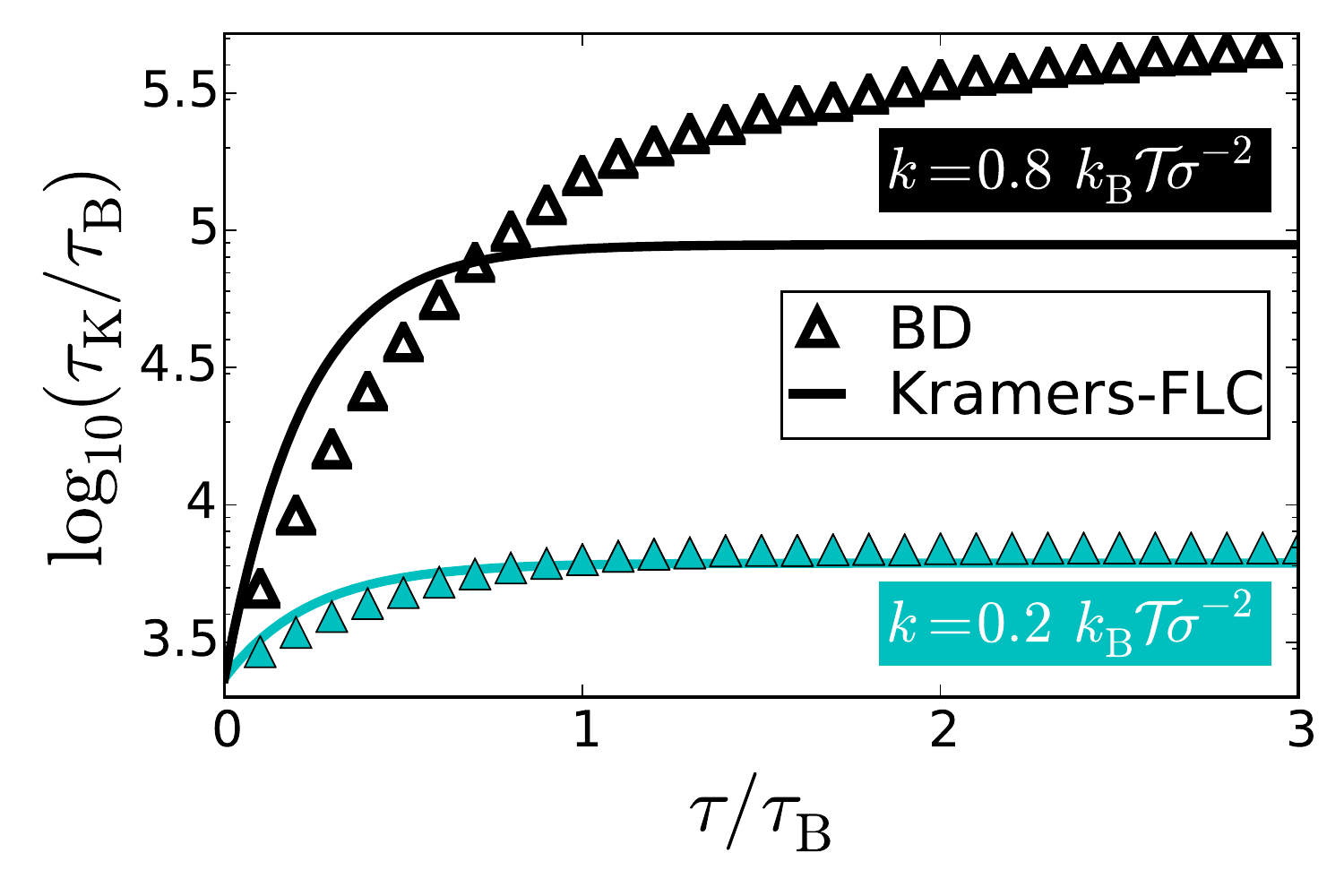}
\caption{(Color online) Logarithm of the mean escape time $\tau_\mathrm{K}$ for the ``washboard'' as a function of the delay time $\tau$. Cyan color: delay force amplitude $k\!=\!0.2\,k_{\mathrm{B}}{\cal T}/\sigma^2$, black color: $k\!=\!0.8\,k_{\mathrm{B}}{\cal T}/\sigma^2$. Symbols: BD, lines: Kramers-FLC estimate.
 }\label{FIG:escapetime-vs-tau}
\end{figure}
\subsubsection{The Kramers-FLC estimate}
The escape time $\tau_\mathrm{K}$ denotes the average time $\chi(t)$ spends in the vicinity of a potential minimum, until it leaves the valley by jumping to an adjacent one. By construction, mean escape times involve two locations and two instances in time, and hence are connected to two-time probabilities. Within our FLC approach, the higher order probabilities $\rho^{\mathrm{FLC}}_\mathrm{(n>2),ss}$ are multivariate Gaussian densities. These belong to a quadratic static potential with only one valley and no escape process. Therefore, there is no direct route to calculate the mean escape time on the basis of the FLC approach.
\par
However, in the limit of small noise intensities, we can find an approximation for $\tau_\mathrm{K}$ by using Kramers theory for (Markovian) systems characterized by a static potential landscape $U$. When the potential barriers $\Delta  U$ are large compared to the thermal energy, i.e., when $\gamma D_0\! \ll \! \Delta  U$, the Kramers theory provides an Arrhenius formula \cite{Gardiner2002,Gernert2014,Hanggi1990} for the escape rate, $r_\mathrm{K}
\!=\!{\sqrt{U''( x_{\mathrm{min}})|U''(x_{\mathrm{max}})|}}/({2\pi \gamma}) \exp \left(-{\Delta  U}/{\gamma D_0} \right)$, (with $U''(x_{\mathrm{ex}})\!=\!\partial_{xx}U( x)|_{x=x_\mathrm{ex}}$, with $x_{\mathrm{ex}}\in \{x_{\mathrm{min}}, x_{\mathrm{max}}  \}$). This estimate for $r_\mathrm{K}$ relies on an equilibrium approximation for $\rho_1$ and does not involve $\rho_2$. Due to the two symmetric ways to leave a valley of the ``washboard'' potential, the resulting mean escape time  can be approximated by $ \tau_\mathrm{K}\! =\! {1}/({2 r_\mathrm{K}})$ \cite{Hanggi1990} and for the double-well potential with the unique direction to exit each valley, by $ \tau_\mathrm{K} = {1}/{r_\mathrm{K}} $.
\par
To use the Kramers theory in the context of the FLC approach, we note that the first member of the FPE hierarchy [Eq.\,(\ref{EQ:dFPE-DL1})] is formally identical to a Markovian (non-delayed) FPE for $\rho_\mathrm{1,ss}$ with a static effective potential $V^\mathrm{FLC}_\mathrm{eff}$ given by Eq.\,(\ref{EQ:Veff}).
Thus, we can directly apply the Arrhenius formula with $U\equiv V^\mathrm{FLC}_\mathrm{eff}$. For the linear delay force in our examples, one just needs to substitute 
$\Delta U\!=\!\Delta \! V_\mathrm{ }+(k/2)   (x_{\mathrm{max}}-x_{\mathrm{min}})$ and $
U''( x_{\mathrm{ex}})=V''_\mathrm{s}(x_{\mathrm{ex}})+k (1-|d^i_2|) $.
Taken altogether, our estimate of $\tau_\mathrm{K}$ involves two separate approximations: the FLC to obtain $V^\mathrm{FLC}_\mathrm{eff}$ and the application of the Arrhenius formula to this non-Markovian system (and implicitly, all simplifications made within the Kramers theory). 
\subsubsection{Escape times in the delayed ``washboard'' potential}
We now apply the Kramers-FLC approximation to calculate the escape times for the delayed ``washboard'' potential. Exemplary results for $\tau_\mathrm{K}$ are plotted in Fig.\,\ref{FIG:escapetime-vs-tau} together with numerical data. For $F_\mathrm{d}\!\equiv\!0$, the BD simulation results and the Kramers-FLC estimate coincide. In the regarded parameter range, $\tau_\mathrm{K}$ generally increases with $k$ and $\tau$, and the FLC results are roughly in agreement with the numerical results. This is rather remarkable, due to the crudeness of applying the Kramers theory to the non-Markovian system. At large values of $\tau$, one observes a qualitatively different behavior: while the escape times resulting from the Kramers-FLC estimate saturate, the corresponding BD data continue to increase with $\tau$. This difference becomes particularly prominent for large $k$. We propose that the discrepancy can be explained as follows. As already discussed in Sec.\,\ref{SEC:washboard}, the conditional probability saturates at a finite value $\tau_\mathrm{sat}$, which is related to the intra-well relaxation time $\tau_\mathrm{ir}$. Accordingly, also $V^\mathrm{FLC}_\mathrm{eff}$, and therewith $\tau_\mathrm{K}$ from the Kramers-FLC approximation, saturate at $\tau_\mathrm{sat}$. In the true non-Markovian system, the escape times not only depend on $\rho_\mathrm{2,ss}$, but also on higher $n$-time probabilities. Therefore they don't necessarily saturate together with $\rho_\mathrm{2,ss}$. Indeed, our numerical investigations reveal that the saturation of $\tau_\mathrm{K}$ sets in at much higher $\tau$ values, see Fig.\,\ref{FIG:escapetime-vs-tau}. For larger $\tau$, another, not yet discussed time scale becomes increasingly important, that is the jump duration time. 
In the present system, this time is roughly $2\,\tau_\mathrm{B}$ (according to BD). For $\tau$ in the range of the jump duration times, the temporal changes of the total potential and the {significant changes of the system state variable  $\chi$ }occur on similar time scales. The interplay between both motions then leads to new (inter-well) dynamical behavior not captured by our approach.
It hence becomes less justified to treat the delayed system as a quasi-Markovian one, whose stochastic (non-delayed) dynamics evolve in a static (effective) potential.
\par
\begin{figure} 
\includegraphics[width=\linewidth]{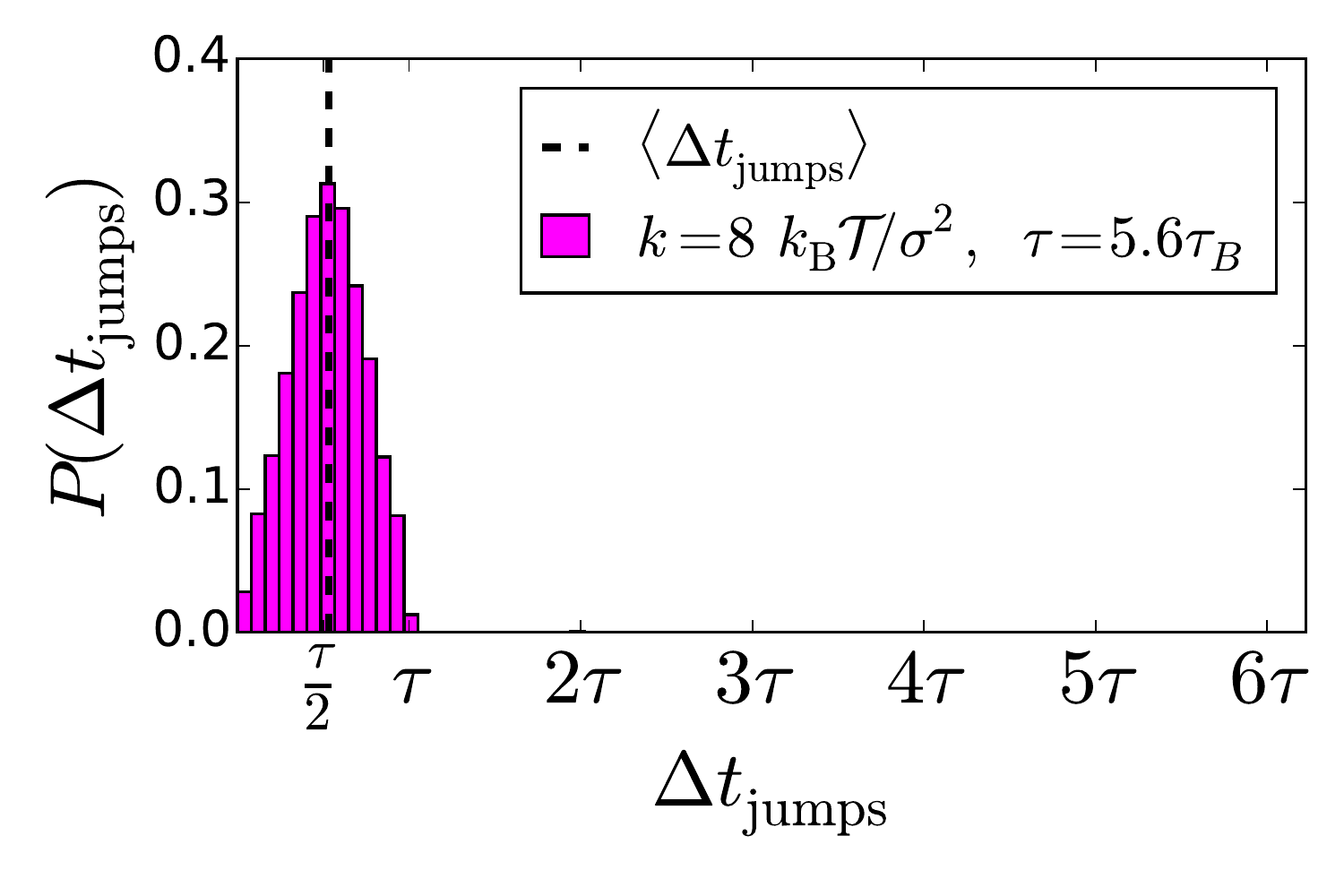}
\caption{(Color online) Waiting time distribution obtained from a numerically (BD) generated (normalized) histogram of time between sequential jumps $\Delta t_\mathrm{jumps}$, for the delayed double-well potential with $k\!=\!8\,k_{\mathrm{B}}{\cal T}/\sigma^2$ and $\tau\!=\!5.6\tau_{\mathrm{B}}$. The dashed black line marks the mean waiting time $\langle\Delta t_\mathrm{jumps} \rangle$.}\label{FIG:waiting-time-dis}
\end{figure}

One example for such new dynamical behavior are quasi-regular oscillations of $\chi$ between two valleys. In fact, we have observed analogous dynamics in the delayed bistable system. For both static potentials, the delay-induced oscillations have a mean period of about $\tau$ and start at random times. We further observe that they occasionally pause and randomly set in again. Figure~\ref{FIG:waiting-time-dis} shows BD results for the distribution of waiting times between sequential jumps in the delayed double-well potential for an exemplary parameter setting. One sees that most waiting times lie within an interval $\Delta t_\mathrm{jump}\in [0,\tau]$ around a single, yet broad peak at about $\tau/2$. This illustrates the stochastic character of the oscillations with mean period of about $\tau$. Similar quasi-regular oscillations can been found in the double-well potential with linear delay force $F_\mathrm{d}(x,x_\tau)\propto x_\tau$ as reported in \cite{Tsimring2001} and further discussed in \cite{Masoller2003}. Interestingly, this version of delayed bistable system has moreover been shown to exhibit coherence resonance, i.e., that the regularity of the oscillations is maximal at a certain finite noise intensity. Preliminary numerical studies suggest that our model systems also display coherence resonance. Further investigations on the delay-induced oscillatory behavior in our systems are in progress.

\section{Spatial autocorrelation function}
\begin{figure}
\includegraphics[width=0.99\linewidth]{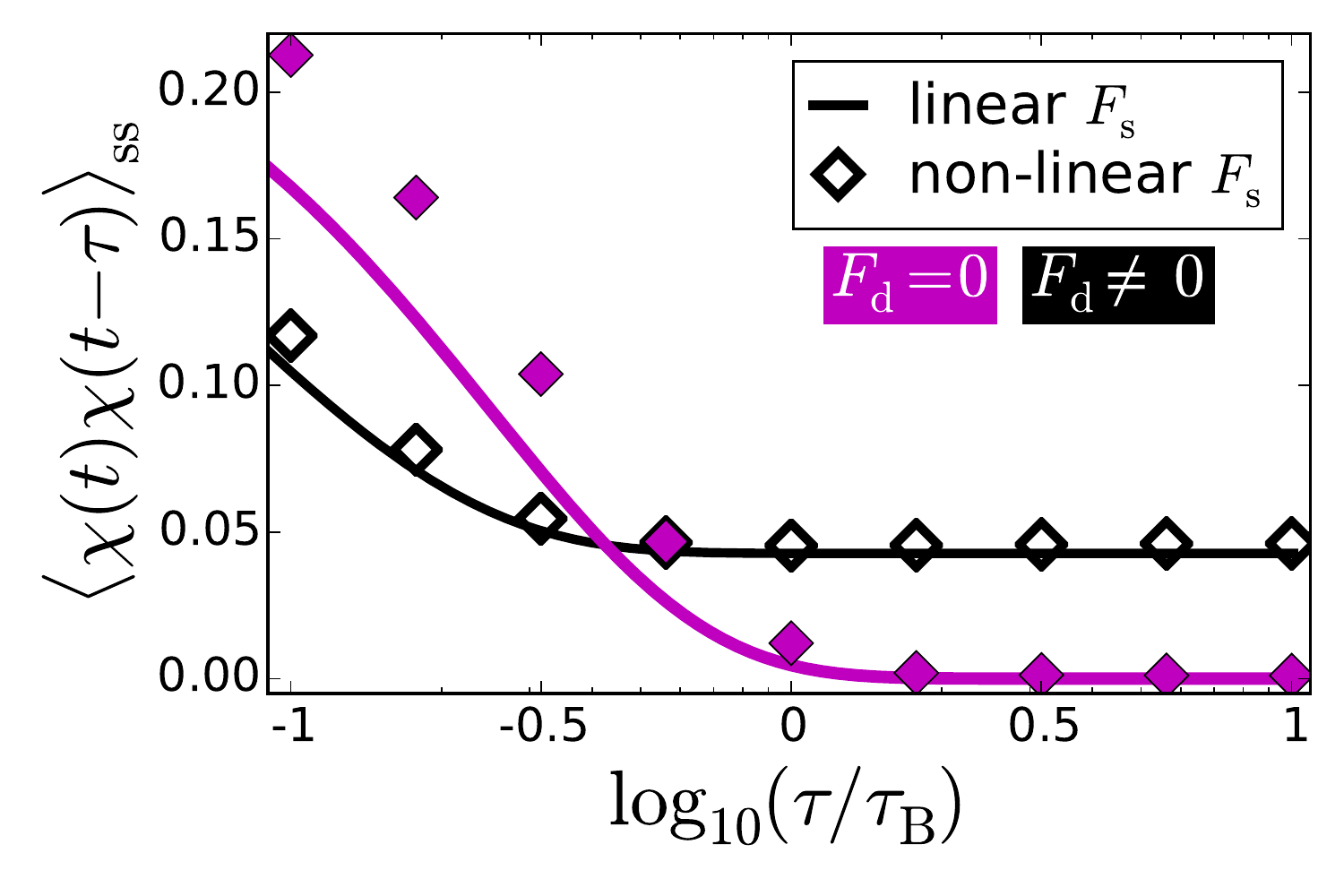}
 \caption{(Color online) Steady state spatial autocorrelation function $C(\tau)\!=\!\langle \chi(t) \chi(t-\tau) \rangle_\mathrm{ss}$ as a function of the delay time $\tau$ within the ``washboard'' potential and the force-linearized version of it. Symbols: BD results for the ``washboard'', solid lines: exact results [Eq.\,(\ref{EQ:Def_C})] for the force linearized systems, black color: $k\!=\!8\, k_{\mathrm{B}}{\cal T} /\sigma^2$, magenta color: $k\!=\!0$.
 }\label{FIG:rho2_correlations}
\end{figure}
In this final section, we aim to discuss a fundamental difference between the FLC and the PT, which has not been in the focus so far.  
This point also provides an explanation why the FLC performs better on the exemplary systems considered in the present work.
\par
We recall that both approaches yield closed FPEs for $\rho_1$ via an approximation of the conditional probability $\rho_\mathrm{c,ss}(x_\tau,t\!-\!\tau|x,t)$, and therewith of the two-time probability density $\rho_\mathrm{2,ss}(x,t;x_\tau,t\!-\!\tau)$ of the delayed system. The essential difference is that in the framework of the FLC, $\rho_\mathrm{c,ss}$ is approximated by the corresponding function of the linearized delayed system, while in the PT approach, $\rho_\mathrm{c,ss}$ stems from the corresponding non-delayed system. The FLC hence involves a $\rho_2$ related to a non-Markovian system, while the PT utilizes a ``Markovian $\rho_2$''. On the level of the two-time probability density, the difference between a Markovian and a non-Markovian system manifests itself, e.g., via the spatial autocorrelation $C(\tau)=\langle \chi(t) \chi(t-\tau) \rangle_\mathrm{ss}-\langle \chi(t) \rangle^2_\mathrm{ss}$ obtained by integrating $\rho_2$ times $x x_\tau$ over both spatial variables and subtracting the squared first moment, see Eq.\,(\ref{EQ:Def_C}). While in the Markovian case $C(\tau)$ always vanishes for large $\tau$, reflecting the loss of memory, it stays finite for typical non-Markovian systems, revealing that the system states at the times $t$ and $t-\tau$ are correlated for large $\tau$. This is illustrated in Fig.\,\ref{FIG:rho2_correlations}, where we have plotted $C(\tau)$ for two of the considered non-Markovian systems (the ``washboard'' potential and the quadratic potential received by the force linearization, both with linear delay force), and for the corresponding Markovian systems (obtained by setting the delay force zero). 
The fact that only the FLC involves a non-Markovian quantity to approximate $\rho_\mathrm{c,ss}$, which accounts for these finite correlations, gives an explanation why it yields better results for large delay times than the perturbation theory.
\par
\section{Conclusion\label{SEC:Conclusion}}
In this work, we introduce the FLC as a novel approach to close the FPE for the steady state probability density of delayed systems.
The new closure is achieved by linearizing the total deterministic force 
in all members of the FPE hierarchy starting from the second. {We have specified on a linear delay force, which can experimentally be implemented, for instance, by an optical laser tweezer acting on a polarizable colloidal particle \cite{Kotar2010}. However, the novel approach generally also applies to nonlinear delay forces}. 
\par 
By considering two exemplary model systems, we present numerical evidence that the novel approach generates indeed a very good approximation of the one-time probability density. In fact, the FLC performs much better here compared to the small delay (Taylor) expansion introduced in Ref.\,\cite{Guillouzic1999}, and to the perturbation theory on the level of the FPE from Ref.\,\cite{Frank2005}.
We also note that the FLC is somewhat more general than the PT in so far as it does not rely (contrary to the PT) on an analytical expression for the conditional probability for the corresponding system without delay force. And even\textit{ if} this quantity is available, the respective quantity of the corresponding \textit{linearized}, delayed system $\rho^\mathrm{lin}_\mathrm{c,ss}$ (which is the key ingredient within the FLC), often renders a better approximation, especially for large delay times. This is because $\rho^\mathrm{lin}_\mathrm{c,ss}$ is a \textit{non-Markovian} quantity which can account for the finite correlations between the system states at the times $t$ and $t\!-\!\tau$. These correlations are not represented in the PT. Moreover, the FLC is not restricted to small noise levels, as the small delay expansion, or to small feedback forces, as the PT approach.
\par
However, the FLC also has its limitations. It only works in the steady state, in contrast to the earlier approaches. Additionally, also the corresponding force linearized system must have steady state. Another issue occurs when our estimate for the total potential in the steady state $V_\mathrm{STAT}$ [Eq.\,(\ref{EQ:V_STAT})] has no minima giving rise to an ambiguity regarding the points around which the linearization should be carried out. However, one can easily adjust the ansatz of the FLC, by simply using other (e.g., random) centers of the linearization, preferring those where the  probability density is high (in order to minimize the error in the final approximation). For instance, the minima of the static potential can be reasonable choices.
\par
Having in mind these restrictions, the FLC can be applied to a much wider class of delayed systems than the ones considered in the present work. One important extension are systems with inertia, for which the Fokker-Planck description has been discussed extensively in Ref.\,\cite{Rosinberg2015}. Such systems have been shown to exhibit complex dynamical behavior including chaos; an example is a delayed double-well potential with inertia and periodic forcing \cite{Nbendjo2003}. 
Furthermore, it is in principle feasible to generalize the FLC towards systems with multiple discrete delays, which, for instance, can play a crucial role in the stochastic process of gene regulation \cite{Parmar2015, *Bratsun2005} or in the collective noisy motion of animals \cite{McKetterick2014}. The force-linearized case of such systems is still solvable \cite{Giuggioli2016, McKetterick2014}, with the $n$-time probabilities being again given by multivariate Gaussian distributions. Also for the more general class of noisy systems with distributed delays, the force linearized case can be handled analytically for arbitrary memory kernels \cite{Giuggioli2016, Hanggi1978,*Hanggi1982,*Hernandez1983, Budini2004, McKetterick2014}. This generally paves the path to adapt the basic idea of the FLC.
However, to the best of our knowledge, the derivation of the FPE with multiple or distributed delays and \textit{nonlinear} forces has not been carried out to date, at least not in the general case. A lack of this equation inherently limits the applicability of the FLC. Another important generalization would be the case of multiplicative noise (and delay), for which so far not many analytical results are known. It is clear that the situation will substantially depend on the specific functional form of the noise term. While at least the delayed Fokker-Planck equation has been worked out for the general case (independent of the specific multiplicative noise term) \cite{Guillouzic1999, Frank2004}, there is to the best of our knowledge no general solution for the linear force case at hand. For some specific multiplicative noise terms, however, the solution for the linear force model is indeed known \cite{Frank2001, Frank2004, Budini2004}, making the FLC in fact feasible (with non-Gaussian conditional probability distributions). Generally speaking, for all types of delayed stochastic systems, the applicability of the FLC depends on the availability of both, the general FPE and the solution of the force-linearized case.
\par
A further aspect touched in the present work concerns the calculation of transport-related quantities such as escape times or waiting time distributions. In general, they require the two-time probability $\rho_2$ or even higher $n$-time densities. In the present form, the FLC does not provide access to $\rho_{n \ge 1}$. Still, we have demonstrated that a reasonable estimate of the escape times is possible by combining the FLC and the Kramers theory. This estimate, however, breaks down when the delay times are in the range of (or large compared to) the jump duration times. Under such conditions, the interplay between the dynamics of the system state variable $\chi$ and the time-dependent energy landscape causes oscillatory motion, whose description is clearly beyond the rather crude FLC-Kramers approximation. For this reason, an extension of the theory towards higher members of the FPE hierarchy would be worthwhile. The analytical results concerning the second member of the FPE hierarchy and $\rho_3$ for the linear case, given in the Appendices \ref{APP:FPE_rho2}-\ref{APP:C}, are essential steps in this direction.
\par
A different way to access transport-related quantities was presented in \cite{Tsimring2001}, where the autocorrelation function in a very similar model (a double-well potential supplemented with linear delay force) was successfully approximated by that of an appropriate discrete model (where an entire potential valley is regarded as one discrete state). In this case the Master equation can analytically be solved. We aim to stress that, in its present form, the FLC does not compete with this approach, but yields a description on a different time and length scale. While Ref.\,\cite{Tsimring2001} yields a convincing approximation of the inter-well dynamics, our approach rather targets the intra-well dynamics. Hence, to some extent, both approaches complement each other. 
\par
Finally, we want to briefly comment on the possible relevance of the FLC for stochastic thermodynamics. Indeed the thermodynamics of delayed systems is a subject of strong current interest \cite{Rosinberg2015,Jiang2011,*Munakata2009,*Munakata2014}. Of particular interest are the information {and entropy change rates}, which rely on two-time probabilities. Time-delayed feedback forces have been shown to yield nontrivial contributions to these quantities \cite{Rosinberg2015, Xiao2016}. In its present form, the framework of the FLC does not provide access to the involved probabilities. This is another motivation to extend our approach towards higher $n$-time probability densities. One could then use the FLC predictions in order to estimate these intriguing thermodynamic quantities far from equilibrium.
\begin{acknowledgments}
This work was supported by the Deutsche Forschungsgemeinschaft through SFB 910 (project B2).
\end{acknowledgments}
%
\appendix
\section{\uppercase{Second member of the delayed FPE, general case}\label{APP:FPE_rho2}}
In the following we present a derivation of the delayed FPE for $\rho_2$, which, to the best of our knowledge, has not been reported elsewhere so far. We extend the derivation presented in Ref.\,\cite{Frank2005} [on which we comment before Eqs.\,(\ref{EQ:dFPE})], towards the second member of the infinite FPE hierarchy. 
Accordingly, we start by considering the formal time derivative of $\rho_2(x,t;x_\tau,t\!-\!\tau)\!=\!\big\langle \delta[x-\chi(t)] \delta[x_\tau-\chi(t\!-\!\tau)] \big\rangle$, and use the delayed LE (\ref{EQ:LE}) to {substitute} the derivatives $ {\partial \chi(t)}/{\partial t}$ and ${\partial\chi(t\!-\!\tau)}/{\partial t}$. We then introduce the functional $\Lambda[\Gamma]\equiv \delta(x-\chi(t))\delta(x_\tau-\chi(t-\tau))$, just like in Ref.\,\cite{Frank2005}. Using Novikov's theorem [see Eq.\,(\ref{EQ:Novikov})], we can express the emerging correlations $\big \langle \Lambda[\Gamma]\Gamma(t)  \big \rangle$ by functional derivatives. We now use basic variational derivative rules:
\begin{equation}\label{Eq:var-derivative_rule}
\frac{\delta \Lambda[\Gamma]}{\delta \Gamma(t)}=   \frac{\delta \Lambda[\Gamma]}{\delta \chi(t)} \frac{\delta \chi(t)}{\delta \Gamma(t)} 
+ \frac{\delta \Lambda[\Gamma]}{\delta \chi(t\!-\!\tau)} \underbrace{\frac{\delta \chi(t\!-\!\tau)}{\delta \Gamma(t)} }_{\rightarrow 0} ,
\end{equation} 
and analogously for ${\delta \Lambda[\Gamma]}/{\delta \Gamma(t-\tau)}$. The last term in Eq.\,(\ref{Eq:var-derivative_rule}) vanishes, since it is noncausal: The particle position at time $t-\tau$ cannot depend on the noise at the later time $t$, since $\Gamma$ is a random force with vanishing temporal correlation \cite{Frank2003}.
In the case of additive noise, the variational derivatives ${\delta \chi(t)}/{\delta \Gamma(t)}$ and ${\delta \chi(t)}/{\delta \Gamma(t-\tau)}$ do not explicitly depend on $\Gamma$, so we can evaluate the ensemble averages and obtain the FPE for $\rho_2$
\begin{align}\label{EQ:General_dFPE2}
&\frac{\partial}{\partial t} \rho_2(x,t;x_\tau,t\!-\!\tau)=  \nonumber \\
& -\frac{\partial}{\partial x} \left\lbrace \frac{F(x,x_\tau) }{\gamma} \rho_2(x,t;x_\tau,t\!-\!\tau) \right\rbrace \nonumber \\
& -\frac{\partial}{\partial x_\tau}\!\! \left\lbrace  \int_{\Omega} \!\frac{F(x_\tau,x_{2\tau}) }{\gamma}\rho_3(x,t;x_\tau,t\!-\!\tau;x_{2\tau},t\!-\!{2\tau}) \mathrm{d}x_{2\tau}\!\right\rbrace \nonumber  \\
& +\!\text{\small $\sqrt{2D_0}$} \,\Bigg( \!\left[ \frac{\partial^2}{\partial x^2} + \frac{\partial^2}{\partial x_\tau^2} \right ] \!\!\left\{ \! \frac{\delta \chi(t)}{\delta \Gamma(t)}\Big|_{\!\!\!\!\!\!\! \!\!\!  \chi(t)=x \atop\!\!  \chi(t\!-\!\tau)=x_\tau}\!\!\!\! \rho_2(x,t;x_\tau,t\!-\!\tau) \! \right\}\nonumber  \\ 
& +\frac{\partial^2}{\partial x_\tau \partial x}    \left\{ \frac{\delta \chi(t)}{\delta \Gamma(t\!-\!\tau)}\Big|_{\!\!\!\!\!\!\! \!\!\!  \chi(t)=x \atop\! \!  \chi(t\!-\!\tau)=x_\tau} \! \!\! \rho_2(x,t;x_\tau,t\!-\!\tau)   \right\}\!\Bigg) .
\end{align}
This is the second member of the Fokker-Planck hierarchy for the general case of a system with delay force $F(x,x_\tau)$. As expected, it contains the next higher-order probability distribution function, that is $\rho_3$. We also note that, in contrast to the first member of the FPE hierarchy [see Eq.\,(\ref{EQ:dFPE})], Eq.\,(\ref{EQ:General_dFPE2}) contains the variational derivatives ${\delta \chi(t)}/{\delta \Gamma(t)}$ and ${\delta \chi(t)}/{\delta \Gamma(t-\tau)}$. These need to be specified for the specific $F$ under consideration.
\section{\uppercase{Second member of FPE for linear forces}\label{APP:FPE_rho2_linear}}
Now we specialize the delayed FPE for $\rho_2$ [see Eq.\,(\ref{EQ:General_dFPE2})] for forces that are linear in $x$ and $x_\tau$, i.e., forces of the form $F^{\mathrm{L}}(x,x_{\tau})=-{\alpha x} -\beta  x_\tau $. The corresponding delayed LE can be solved iteratively by using the method of steps \cite{Frank2003}: to solve the equation in a time interval $t\in[n\tau,(n\!+\!1)\tau], \forall n\in \mathbb{N}$, one substitutes the solution of the preceding interval $[(n\!-\!1)\tau,n\tau]$ to get rid of the delay term (for $n\!=\!0$ the history function is used). One finds that on each interval, the formal solution has the same structure 
\begin{align}\label{EQ:Formal-sol-chi-linear}
\mathcal{Y}_{n+1}(t)=\mathcal{Y}_{n}(n\tau)e^{-\alpha(t-n\tau)/({\gamma\sigma})}+ \nonumber \\
\int_{n\tau}^{t}e^{-\alpha (t-t')/({\gamma\sigma})}[\sqrt{2D_0} \Gamma(t')-\beta \mathcal{Y}_n(t'-\tau)]\mathrm{d}t',
\end{align} 
with $\chi(t)=\mathcal{Y}_{n+1}(t)$, for $t\in [n\tau,(n+1)\tau]$. From Eq.\,(\ref{EQ:Formal-sol-chi-linear}) we find the functional derivatives
\begin{align}
\frac{\delta \chi(t)}{\delta \Gamma(t )}=& \frac{\delta \chi(t\!-\!\tau)}{\delta \Gamma(t\!-\!\tau)}=\sqrt{\frac{D_0}{2}}, \label{EQ:derivative1} \\
\frac{\delta \chi(t)}{\delta \Gamma(t\!-\!\tau)} = &\sqrt{2 D_0} \, e^{{-\alpha \tau}/({\gamma\sigma})}. \label{EQ:derivative2}
\end{align}
Inserting Eqs.\,(\ref{EQ:derivative1}) and (\ref{EQ:derivative2}) into Eq.\,(\ref{EQ:General_dFPE2}) we obtain for the case of linear forces $F^{\mathrm{L}}(x,x_{\tau})=-{\alpha x} -\beta  x_\tau $:
\begin{align}\label{EQ:dFPE2}
&\frac{\partial}{\partial t} \rho_2(x,t;x_\tau,t\!-\!\tau)= \nonumber \\
& - \frac{\partial}{\partial x} \left\lbrace \frac{F^{\mathrm{L}}(x,x_{\tau})}{\gamma} \rho_2(x,t; x_{\tau}, t\!-\!\tau)\right\rbrace  \nonumber \\
& - \frac{\partial}{\partial x_\tau} \int_{\Omega}\!\frac{F^{\mathrm{L}}(x_\tau,x_{2\tau})}{\gamma}  \rho_3(x,t; x_{\tau}, t\!-\!\tau;x_{2\tau},t\!-\!2\tau)\, \mathrm{d}x_{2\tau} \nonumber \\
& + D_0 \left[ \frac{\partial ^2}{\partial x ^2} +\frac{\partial ^2}{\partial x_\tau ^2} \right]\rho_2(x,t;x_\tau,t\!-\!\tau) \nonumber \\
& + 2 D_0 e^{{-\alpha \tau}/({\gamma\sigma})} \frac{\partial ^2}{\partial x\partial x_\tau} \rho_2(x,t;x_\tau,t\!-\!\tau).
\end{align} 
In the steady state, Eqs.\,(\ref{EQ:dFPE2}) and the FPE for $\rho_1$ [Eq.\,(\ref{EQ:dFPE})] are solved by the respective (multivariate) Gaussian distributions \cite{Kuechler1992}
\begin{equation}\label{EQ:rho_n-linear}
\rho_{\mathrm{n, ss}} 
=\frac{1}{\sqrt{(2\pi)^n \det \{\mathbf{D_{n}}\} }}e^{-{(1/2)}(\bm{x}-\langle \bm{x} \rangle) \mathbf{D_{n}}^{-1} (\bm{x}-\langle \bm{x}\rangle)},
\end{equation}
with $\bm{x}\!=\!(x,x_\tau,..,x_{(n-1)\tau})^\mathrm{T}$, $\mathbf{D_{1}}\!=\! C(0)$, and
\begin{small}
\begin{equation} 
\mathbf{D_{2}}= 
\begin{pmatrix}
C(0)& C(\tau)\\
C(\tau)& C(0)\\
\end{pmatrix},
\mathbf{D_{3}}= 
\begin{pmatrix}
C(0)& C(\tau)& C(2\tau)\\
C(\tau)& C(0)& C(\tau)\\
 C(2\tau)& C(\tau)& C(0)\\
\end{pmatrix}.
\end{equation}
\end{small}
The covariance matrices $\mathbf{D_n}$ involve the steady state spatial autocorrelation function between the systems state at time $t$ and time $t+z$:
\begin{equation}\label{EQ:Def_C}
C(z):=\langle \chi(t)\chi(t+z) \rangle_{\mathrm{ss}}  -\langle \chi(t)\rangle^2_{\mathrm{ss}}  .
\end{equation}
Explicit expressions for $C(0)$, $C(\tau)$, and $C(2\tau)$ (which occur in $\mathbf{D_{k}}$, k$=1,2,3$) are given in the next section.
\section{SPATIAL AUTOCORRELATION FUNCTION\label{APP:C}}
Due to the symmetry property $C(z)=C(-z)$, we only consider non-negative arguments $z$ in the following.
As shown in Refs.\,\cite{Guillouzic1999, Frank2003}, the spatial autocorrelation function on the interval $z\!\in \! [0,\tau]$ is given by
\begin{align}\label{EQ:Corr-0}
C(0)=& D_0 \frac{\gamma+(\beta /\omega) \sinh(\omega \tau)}{\alpha+\beta \cosh(\omega \tau)}, \\
C(z)=& C(0)\cosh(\omega z) - ({D_0}/\omega) \sinh(\omega |z|)\label{EQ:Corr-tau}
\end{align}
with $\omega\!=\!\sqrt{\!\alpha^2-\beta^2}/\gamma \in\! \mathbb{C}$. In order to obtain $C(z)$ for $z> 0$, one can directly deduce from the delayed LE \cite{Frank2003}:
\begin{align}\label{EQ:DGL_C-1}
&\frac{\mathrm{d}C(z)}{\mathrm{d}z}=\Bigg \langle \chi(t) \frac{\mathrm{d}\chi(u)}{\mathrm{d}u}\Big|_{u=t+z} \Bigg\rangle_\mathrm{ss}\nonumber \\
&=-\frac{\alpha}{\gamma} C(z) -\frac{\beta}{\gamma} C(z-\tau) +\sqrt{2D_0} \langle \chi(t)\Gamma(t+z) \rangle.
\end{align}
For $z>0$, the last term vanishes, since a non-zero correlation between the system state $\chi$ and the future noise would be noncausal. The resulting differential equation for $z>0$
\begin{align}\label{EQ:DGL_C}
&{\mathrm{d}C(z)}/{\mathrm{d}z}=-({\alpha}/{\gamma}) C(z) -({\beta}/{\gamma}) C(z-\tau),
\end{align}
 can be solved iteratively using the method of steps (see text at the beginning of the Appendix \ref{APP:FPE_rho2_linear}). Using the solution for the interval $z\in [0,\tau]$ in Eqs.\,(\ref{EQ:Corr-0}) and (\ref{EQ:Corr-tau}), we obtain the correlation function on the interval $z\in [\tau, 2\tau]$,
\begin{align}\label{EQ:Corr-2tau}
C(z)=& D_0 
\frac{2  e^{\alpha (\tau -|z|)} [\beta  \cosh( \omega \tau) + \alpha ]+[...]^*}{2 \beta [\alpha + \beta \cosh(\omega\tau)]},
\end{align} 
where
\begin{align}\label{EQ:Corr-2tau_b}
[...]^*=& - 2 \alpha  \cosh[\omega (\tau-z)]
- \beta  \cosh[\omega (2\tau-z)] \nonumber \\
& +[(\beta^2\!-\!2 \alpha^2)/\omega] \sinh[\omega (\tau -|z|)] 
\nonumber \\
&- (\alpha\beta/\omega) \sinh[\omega ( 2 \tau-|z|)].
\end{align} 
Inspection of Eq.\,(\ref{EQ:Corr-2tau}) shows that there exist critical $\tau$-values for which $C(z)$ diverges, see also text after Eq.\,(\ref{EQ:d2}) (with $\alpha\!=\!\alpha_i$ and $\beta\!=\!\beta_i$ in the discussion there).
\par
We would like to note that in Ref.\,\cite{Frank2003}, the solution (\ref{EQ:Corr-tau}) is erroneously stated to be valid for all $z\!\in\!\mathbb{R}$. Along their derivation, Eq.\,(18) of Ref.\,\cite{Frank2003} is actually not valid for $|z|\!>\!\tau$, although they claim otherwise. A correlation function (\ref{EQ:Corr-tau}) $\forall z\!\in\!\mathbb{R}$ would in fact be unphysical. Consider for example $\beta\!<\!\alpha$, where the system is spatially confined. Then, it is not reasonable that the autocorrelation function grows without bounds for large time differences, as Eq.\,(\ref{EQ:Corr-tau}) would predict. The iterative solution obtained from Eq.\,(\ref{EQ:DGL_C}) by using the method of steps, is indeed bounded, consisting with the physical intuition.
\section{BROWNIAN DYNAMICS SIMULATIONS\label{APP:BD}}
In order to test our theoretical results against quasi-exact data, we perform Brownian dynamics (BD) simulations of the delayed LE (\ref{EQ:LE}). We use the Euler-Maruyama integration scheme \cite{Buckwar2000, Kloeden1992}, with a varying temporal discretization of $\Delta t\!\in \! [ 10^{-3},10^{-6} ]\,\tau_\mathrm{B}$, such that $ 1000\,\Delta t\le \tau$. We perform each simulation multiple times (with different random number seed), and take the average over at least  $10^5$ realizations. {The (pseudo) random numbers are generated with the algorithm ``Mersenne Twister'' \cite{Mersenne} and the Box-M\"uller method \cite{Kloeden1992}.
\par
As initial condition, we use an equilibrium configuration in the respective static potential without delay force.} Before we measure the steady state properties, we let $100\,\tau_\mathrm{B}$ times pass in the presence of the delay force to let the system reach a steady state. The simulation time thereafter is more than $200\,\tau_\mathrm{B}$.
\par
The density profiles from the BD simulations are obtained from histograms with a spatial resolution of $\Delta x \!=\! 0.005\,\sigma$. These histograms are also used to calculate the moments $\mu_n$. With the same procedure the moments of the analytical distributions resulting from the FLC, PT and small delay expansion are calculated (we use the same bin sizes and positions as in the simulations). 
%
%
%
%
%

\begin{thebibliography}{80}%
\makeatletter
\providecommand \@ifxundefined [1]{%
 \@ifx{#1\undefined}
}%
\providecommand \@ifnum [1]{%
 \ifnum #1\expandafter \@firstoftwo
 \else \expandafter \@secondoftwo
 \fi
}%
\providecommand \@ifx [1]{%
 \ifx #1\expandafter \@firstoftwo
 \else \expandafter \@secondoftwo
 \fi
}%
\providecommand \natexlab [1]{#1}%
\providecommand \enquote  [1]{``#1''}%
\providecommand \bibnamefont  [1]{#1}%
\providecommand \bibfnamefont [1]{#1}%
\providecommand \citenamefont [1]{#1}%
\providecommand \href@noop [0]{\@secondoftwo}%
\providecommand \href [0]{\begingroup \@sanitize@url \@href}%
\providecommand \@href[1]{\@@startlink{#1}\@@href}%
\providecommand \@@href[1]{\endgroup#1\@@endlink}%
\providecommand \@sanitize@url [0]{\catcode `\\12\catcode `\$12\catcode
  `\&12\catcode `\#12\catcode `\^12\catcode `\_12\catcode `\%12\relax}%
\providecommand \@@startlink[1]{}%
\providecommand \@@endlink[0]{}%
\providecommand \url  [0]{\begingroup\@sanitize@url \@url }%
\providecommand \@url [1]{\endgroup\@href {#1}{\urlprefix }}%
\providecommand \urlprefix  [0]{URL }%
\providecommand \Eprint [0]{\href }%
\providecommand \doibase [0]{http://dx.doi.org/}%
\providecommand \selectlanguage [0]{\@gobble}%
\providecommand \bibinfo  [0]{\@secondoftwo}%
\providecommand \bibfield  [0]{\@secondoftwo}%
\providecommand \translation [1]{[#1]}%
\providecommand \BibitemOpen [0]{}%
\providecommand \bibitemStop [0]{}%
\providecommand \bibitemNoStop [0]{.\EOS\space}%
\providecommand \EOS [0]{\spacefactor3000\relax}%
\providecommand \BibitemShut  [1]{\csname bibitem#1\endcsname}%
\let\auto@bib@innerbib\@empty
\bibitem [{\citenamefont {Longtin}(2010)}]{Longtin2010}%
  \BibitemOpen
  \bibfield  {author} {\bibinfo {author} {\bibfnamefont {A.}~\bibnamefont
  {Longtin}},\ }\href@noop {} {\emph {\bibinfo {title} {Complex time-delay
  systems: theory and applications}}},\ edited by\ \bibinfo {editor}
  {\bibfnamefont {F.~M.}\ \bibnamefont {Atay}}\ (\bibinfo  {publisher}
  {Springer-Verlag Berlin Heidelberg},\ \bibinfo {year} {2010})\ pp.\ \bibinfo
  {pages} {177--195}\BibitemShut {NoStop}%
\bibitem [{\citenamefont {Sch{\"o}ll}\ \emph {et~al.}(2016)\citenamefont
  {Sch{\"o}ll}, \citenamefont {Klapp},\ and\ \citenamefont
  {H{\"o}vel}}]{Schoell2016}%
  \BibitemOpen
  \bibinfo {editor} {\bibfnamefont {E.}~\bibnamefont {Sch{\"o}ll}}, \bibinfo
  {editor} {\bibfnamefont {S.~H.~L.}\ \bibnamefont {Klapp}}, \ and\ \bibinfo
  {editor} {\bibfnamefont {P.}~\bibnamefont {H{\"o}vel}},\ eds.,\ \href@noop {}
  {\emph {\bibinfo {title} {Control of self-organizing nonlinear systems}}}\
  (\bibinfo  {publisher} {Springer},\ \bibinfo {year} {2016})\BibitemShut
  {NoStop}%
\bibitem [{\citenamefont {Sch{\"o}ll}\ and\ \citenamefont
  {Schuster}(2008)}]{Schoell2008}%
  \BibitemOpen
  \bibinfo {editor} {\bibfnamefont {E.}~\bibnamefont {Sch{\"o}ll}}\ and\
  \bibinfo {editor} {\bibfnamefont {H.~G.}\ \bibnamefont {Schuster}},\ eds.,\
  \href@noop {} {\emph {\bibinfo {title} {Handbook of chaos control}}}\
  (\bibinfo  {publisher} {John Wiley \& Sons},\ \bibinfo {year}
  {2008})\BibitemShut {NoStop}%
\bibitem [{\citenamefont {Montinaro}\ \emph {et~al.}(2012)\citenamefont
  {Montinaro}, \citenamefont {Mehlin}, \citenamefont {Solanki}, \citenamefont
  {Peddibhotla}, \citenamefont {Mack}, \citenamefont {Awschalom},\ and\
  \citenamefont {Poggio}}]{Montinaro2012}%
  \BibitemOpen
  \bibfield  {author} {\bibinfo {author} {\bibfnamefont {M.}~\bibnamefont
  {Montinaro}}, \bibinfo {author} {\bibfnamefont {A.}~\bibnamefont {Mehlin}},
  \bibinfo {author} {\bibfnamefont {H.}~\bibnamefont {Solanki}}, \bibinfo
  {author} {\bibfnamefont {P.}~\bibnamefont {Peddibhotla}}, \bibinfo {author}
  {\bibfnamefont {S.}~\bibnamefont {Mack}}, \bibinfo {author} {\bibfnamefont
  {D.}~\bibnamefont {Awschalom}}, \ and\ \bibinfo {author} {\bibfnamefont
  {M.}~\bibnamefont {Poggio}},\ }\href@noop {} {\bibfield  {journal} {\bibinfo
  {journal} {Appl. Phys. Lett.}\ }\textbf {\bibinfo {volume} {101}},\ \bibinfo
  {pages} {133104} (\bibinfo {year} {2012})}\BibitemShut {NoStop}%
\bibitem [{\citenamefont {Stoica}(2005)}]{Stoica2004}%
  \BibitemOpen
  \bibfield  {author} {\bibinfo {author} {\bibfnamefont {G.}~\bibnamefont
  {Stoica}},\ }\href@noop {} {\bibfield  {journal} {\bibinfo  {journal} {Proc.
  Am. Math. Soc.}\ }\textbf {\bibinfo {volume} {133}},\ \bibinfo {pages} {1837}
  (\bibinfo {year} {2005})}\BibitemShut {NoStop}%
\bibitem [{\citenamefont {Voss}\ and\ \citenamefont {Kurths}(2002)}]{Voss2002}%
  \BibitemOpen
  \bibfield  {author} {\bibinfo {author} {\bibfnamefont {H.~U.}\ \bibnamefont
  {Voss}}\ and\ \bibinfo {author} {\bibfnamefont {J.}~\bibnamefont {Kurths}},\
  }\href@noop {} {\emph {\bibinfo {title} {Modelling and forecasting financial
  data: techniques of nonlinear dynamics}}},\ edited by\ \bibinfo {editor}
  {\bibfnamefont {A.~S.}\ \bibnamefont {Soofi}}\ and\ \bibinfo {editor}
  {\bibfnamefont {L.}~\bibnamefont {Cao}},\ Vol.~\bibinfo {volume} {2}\
  (\bibinfo  {publisher} {Springer Science \& Business Media},\ \bibinfo {year}
  {2002})\ pp.\ \bibinfo {pages} {327--349}\BibitemShut {NoStop}%
\bibitem [{\citenamefont {Longtin}\ \emph {et~al.}(1990)\citenamefont
  {Longtin}, \citenamefont {Milton}, \citenamefont {Bos},\ and\ \citenamefont
  {Mackey}}]{Longtin1990}%
  \BibitemOpen
  \bibfield  {author} {\bibinfo {author} {\bibfnamefont {A.}~\bibnamefont
  {Longtin}}, \bibinfo {author} {\bibfnamefont {J.~G.}\ \bibnamefont {Milton}},
  \bibinfo {author} {\bibfnamefont {J.~E.}\ \bibnamefont {Bos}}, \ and\
  \bibinfo {author} {\bibfnamefont {M.~C.}\ \bibnamefont {Mackey}},\
  }\href@noop {} {\bibfield  {journal} {\bibinfo  {journal} {Phys. Rev. A}\
  }\textbf {\bibinfo {volume} {41}},\ \bibinfo {pages} {6992} (\bibinfo {year}
  {1990})}\BibitemShut {NoStop}%
\bibitem [{\citenamefont {Milton}\ \emph {et~al.}(2013)\citenamefont {Milton},
  \citenamefont {Fuerte}, \citenamefont {B{\'e}lair}, \citenamefont {Lippai},
  \citenamefont {Kamimura},\ and\ \citenamefont {Ohira}}]{Milton2013}%
  \BibitemOpen
  \bibfield  {author} {\bibinfo {author} {\bibfnamefont {J.~G.}\ \bibnamefont
  {Milton}}, \bibinfo {author} {\bibfnamefont {A.}~\bibnamefont {Fuerte}},
  \bibinfo {author} {\bibfnamefont {C.}~\bibnamefont {B{\'e}lair}}, \bibinfo
  {author} {\bibfnamefont {J.}~\bibnamefont {Lippai}}, \bibinfo {author}
  {\bibfnamefont {A.}~\bibnamefont {Kamimura}}, \ and\ \bibinfo {author}
  {\bibfnamefont {T.}~\bibnamefont {Ohira}},\ }\href@noop {} {\bibfield
  {journal} {\bibinfo  {journal} {Nonlinear Theory and Its Applications,
  IEICE}\ }\textbf {\bibinfo {volume} {4}},\ \bibinfo {pages} {129} (\bibinfo
  {year} {2013})}\BibitemShut {NoStop}%
\bibitem [{\citenamefont {Milton}\ \emph {et~al.}(2009)\citenamefont {Milton},
  \citenamefont {Ohira}, \citenamefont {Cabrera}, \citenamefont {Fraiser},
  \citenamefont {Gyorffy}, \citenamefont {Ruiz}, \citenamefont {Strauss},
  \citenamefont {Balch}, \citenamefont {Marin},\ and\ \citenamefont
  {Alexander}}]{Milton2009a}%
  \BibitemOpen
  \bibfield  {author} {\bibinfo {author} {\bibfnamefont {J.~G.}\ \bibnamefont
  {Milton}}, \bibinfo {author} {\bibfnamefont {T.}~\bibnamefont {Ohira}},
  \bibinfo {author} {\bibfnamefont {J.~L.}\ \bibnamefont {Cabrera}}, \bibinfo
  {author} {\bibfnamefont {R.~M.}\ \bibnamefont {Fraiser}}, \bibinfo {author}
  {\bibfnamefont {J.~B.}\ \bibnamefont {Gyorffy}}, \bibinfo {author}
  {\bibfnamefont {F.~K.}\ \bibnamefont {Ruiz}}, \bibinfo {author}
  {\bibfnamefont {M.~A.}\ \bibnamefont {Strauss}}, \bibinfo {author}
  {\bibfnamefont {E.~C.}\ \bibnamefont {Balch}}, \bibinfo {author}
  {\bibfnamefont {P.~J.}\ \bibnamefont {Marin}}, \ and\ \bibinfo {author}
  {\bibfnamefont {J.~L.}\ \bibnamefont {Alexander}},\ }\href
  {http://dx.doi.org/10.1371/journal.pone.0007427} {\bibfield  {journal}
  {\bibinfo  {journal} {PLoS ONE}\ }\textbf {\bibinfo {volume} {4}},\ \bibinfo
  {pages} {e7427} (\bibinfo {year} {2009})}\BibitemShut {NoStop}%
\bibitem [{\citenamefont {Goel}\ \emph {et~al.}(1971)\citenamefont {Goel},
  \citenamefont {Maitra},\ and\ \citenamefont {Montroll}}]{Goel1971}%
  \BibitemOpen
  \bibfield  {author} {\bibinfo {author} {\bibfnamefont {N.~S.}\ \bibnamefont
  {Goel}}, \bibinfo {author} {\bibfnamefont {S.~C.}\ \bibnamefont {Maitra}}, \
  and\ \bibinfo {author} {\bibfnamefont {E.~W.}\ \bibnamefont {Montroll}},\
  }\href@noop {} {\bibfield  {journal} {\bibinfo  {journal} {Rev. Mod. Phys.}\
  }\textbf {\bibinfo {volume} {43}},\ \bibinfo {pages} {231} (\bibinfo {year}
  {1971})}\BibitemShut {NoStop}%
\bibitem [{\citenamefont {Das}\ \emph {et~al.}(2012)\citenamefont {Das},
  \citenamefont {Srinivas}, \citenamefont {Srinivas},\ and\ \citenamefont
  {Gazi}}]{Das2012}%
  \BibitemOpen
  \bibfield  {author} {\bibinfo {author} {\bibfnamefont {K.}~\bibnamefont
  {Das}}, \bibinfo {author} {\bibfnamefont {M.}~\bibnamefont {Srinivas}},
  \bibinfo {author} {\bibfnamefont {M.}~\bibnamefont {Srinivas}}, \ and\
  \bibinfo {author} {\bibfnamefont {N.}~\bibnamefont {Gazi}},\ }\href
  {http://dx.doi.org/10.1016/j.crvi.2012.06.001} {\bibfield  {journal}
  {\bibinfo  {journal} {C. R. Biol.}\ }\textbf {\bibinfo {volume} {335}},\
  \bibinfo {pages} {503–513} (\bibinfo {year} {2012})}\BibitemShut {NoStop}%
\bibitem [{\citenamefont {Parmar}\ \emph {et~al.}(2015)\citenamefont {Parmar},
  \citenamefont {Blyuss}, \citenamefont {Kyrychko},\ and\ \citenamefont
  {Hogan}}]{Parmar2015}%
  \BibitemOpen
  \bibfield  {author} {\bibinfo {author} {\bibfnamefont {K.}~\bibnamefont
  {Parmar}}, \bibinfo {author} {\bibfnamefont {K.~B.}\ \bibnamefont {Blyuss}},
  \bibinfo {author} {\bibfnamefont {Y.~N.}\ \bibnamefont {Kyrychko}}, \ and\
  \bibinfo {author} {\bibfnamefont {S.~J.}\ \bibnamefont {Hogan}},\ }\href@noop
  {} {\bibfield  {journal} {\bibinfo  {journal} {Comput. Math. Meth. M.}\
  }\textbf {\bibinfo {volume} {2015}},\ \bibinfo {pages} {347273} (\bibinfo
  {year} {2015})}\BibitemShut {NoStop}%
\bibitem [{\citenamefont {Bratsun}\ \emph {et~al.}(2005)\citenamefont
  {Bratsun}, \citenamefont {Volfson}, \citenamefont {Tsimring},\ and\
  \citenamefont {Hasty}}]{Bratsun2005}%
  \BibitemOpen
  \bibfield  {author} {\bibinfo {author} {\bibfnamefont {D.}~\bibnamefont
  {Bratsun}}, \bibinfo {author} {\bibfnamefont {D.}~\bibnamefont {Volfson}},
  \bibinfo {author} {\bibfnamefont {L.~S.}\ \bibnamefont {Tsimring}}, \ and\
  \bibinfo {author} {\bibfnamefont {J.}~\bibnamefont {Hasty}},\ }\href@noop {}
  {\bibfield  {journal} {\bibinfo  {journal} {Proc. Natl. Acad. Sci. U. S. A.}\
  }\textbf {\bibinfo {volume} {102}},\ \bibinfo {pages} {14593} (\bibinfo
  {year} {2005})}\BibitemShut {NoStop}%
\bibitem [{\citenamefont {Gupta}\ \emph {et~al.}(2013)\citenamefont {Gupta},
  \citenamefont {L\'opez}, \citenamefont {Ott}, \citenamefont {Josi\'{c}},\
  and\ \citenamefont {Bennett}}]{Gupta2013}%
  \BibitemOpen
  \bibfield  {author} {\bibinfo {author} {\bibfnamefont {C.}~\bibnamefont
  {Gupta}}, \bibinfo {author} {\bibfnamefont {J.~M.}\ \bibnamefont {L\'opez}},
  \bibinfo {author} {\bibfnamefont {W.}~\bibnamefont {Ott}}, \bibinfo {author}
  {\bibfnamefont {K.}~\bibnamefont {Josi\'{c}}}, \ and\ \bibinfo {author}
  {\bibfnamefont {M.~R.}\ \bibnamefont {Bennett}},\ }\href {\doibase
  10.1103/PhysRevLett.111.058104} {\bibfield  {journal} {\bibinfo  {journal}
  {Phys. Rev. Lett.}\ }\textbf {\bibinfo {volume} {111}},\ \bibinfo {pages}
  {058104} (\bibinfo {year} {2013})}\BibitemShut {NoStop}%
\bibitem [{\citenamefont {Masoller}(2002)}]{Masoller2002}%
  \BibitemOpen
  \bibfield  {author} {\bibinfo {author} {\bibfnamefont {C.}~\bibnamefont
  {Masoller}},\ }\href {\doibase 10.1103/PhysRevLett.88.034102} {\bibfield
  {journal} {\bibinfo  {journal} {Phys. Rev. Lett.}\ }\textbf {\bibinfo
  {volume} {88}},\ \bibinfo {pages} {034102} (\bibinfo {year}
  {2002})}\BibitemShut {NoStop}%
\bibitem [{\citenamefont {Hein}\ \emph {et~al.}(2015)\citenamefont {Hein},
  \citenamefont {Schulze}, \citenamefont {Carmele},\ and\ \citenamefont
  {Knorr}}]{Hein2015}%
  \BibitemOpen
  \bibfield  {author} {\bibinfo {author} {\bibfnamefont {S.~M.}\ \bibnamefont
  {Hein}}, \bibinfo {author} {\bibfnamefont {F.}~\bibnamefont {Schulze}},
  \bibinfo {author} {\bibfnamefont {A.}~\bibnamefont {Carmele}}, \ and\
  \bibinfo {author} {\bibfnamefont {A.}~\bibnamefont {Knorr}},\ }\href
  {\doibase 10.1103/PhysRevA.91.052321} {\bibfield  {journal} {\bibinfo
  {journal} {Phys. Rev. A}\ }\textbf {\bibinfo {volume} {91}},\ \bibinfo
  {pages} {052321} (\bibinfo {year} {2015})}\BibitemShut {NoStop}%
\bibitem [{\citenamefont {Lu}\ \emph {et~al.}(2017)\citenamefont {Lu},
  \citenamefont {Naumann}, \citenamefont {Cerrillo}, \citenamefont {Zhao},
  \citenamefont {Knorr},\ and\ \citenamefont {Carmele}}]{Lu2017}%
  \BibitemOpen
  \bibfield  {author} {\bibinfo {author} {\bibfnamefont {Y.}~\bibnamefont
  {Lu}}, \bibinfo {author} {\bibfnamefont {N.~L.}\ \bibnamefont {Naumann}},
  \bibinfo {author} {\bibfnamefont {J.}~\bibnamefont {Cerrillo}}, \bibinfo
  {author} {\bibfnamefont {Q.}~\bibnamefont {Zhao}}, \bibinfo {author}
  {\bibfnamefont {A.}~\bibnamefont {Knorr}}, \ and\ \bibinfo {author}
  {\bibfnamefont {A.}~\bibnamefont {Carmele}},\ }\href@noop {} {\bibfield
  {journal} {\bibinfo  {journal} {arXiv preprint arXiv:1703.10028}\ } (\bibinfo
  {year} {2017})}\BibitemShut {NoStop}%
\bibitem [{\citenamefont {Braun}\ and\ \citenamefont
  {Cichos}(2013)}]{Braun2013}%
  \BibitemOpen
  \bibfield  {author} {\bibinfo {author} {\bibfnamefont {M.}~\bibnamefont
  {Braun}}\ and\ \bibinfo {author} {\bibfnamefont {F.}~\bibnamefont {Cichos}},\
  }\href {http://dx.doi.org/10.1021/nn404980k} {\bibfield  {journal} {\bibinfo
  {journal} {ACS Nano}\ }\textbf {\bibinfo {volume} {7}},\ \bibinfo {pages}
  {11200–11208} (\bibinfo {year} {2013})}\BibitemShut {NoStop}%
\bibitem [{\citenamefont {Braun}\ \emph {et~al.}(2015)\citenamefont {Braun},
  \citenamefont {Bregulla}, \citenamefont {Günther}, \citenamefont {Mertig},\
  and\ \citenamefont {Cichos}}]{Braun2015}%
  \BibitemOpen
  \bibfield  {author} {\bibinfo {author} {\bibfnamefont {M.}~\bibnamefont
  {Braun}}, \bibinfo {author} {\bibfnamefont {A.~P.}\ \bibnamefont {Bregulla}},
  \bibinfo {author} {\bibfnamefont {K.}~\bibnamefont {Günther}}, \bibinfo
  {author} {\bibfnamefont {M.}~\bibnamefont {Mertig}}, \ and\ \bibinfo {author}
  {\bibfnamefont {F.}~\bibnamefont {Cichos}},\ }\href
  {http://dx.doi.org/10.1021/acs.nanolett.5b01999} {\bibfield  {journal}
  {\bibinfo  {journal} {Nano Lett.}\ }\textbf {\bibinfo {volume} {15}},\
  \bibinfo {pages} {5499–5505} (\bibinfo {year} {2015})}\BibitemShut
  {NoStop}%
\bibitem [{\citenamefont {Haeufle}\ \emph {et~al.}(2016)\citenamefont
  {Haeufle}, \citenamefont {Bauerle}, \citenamefont {Steiner}, \citenamefont
  {Bremicker}, \citenamefont {Schmitt},\ and\ \citenamefont
  {Bechinger}}]{Haeufle2016}%
  \BibitemOpen
  \bibfield  {author} {\bibinfo {author} {\bibfnamefont {D.~F.~B.}\
  \bibnamefont {Haeufle}}, \bibinfo {author} {\bibfnamefont {T.}~\bibnamefont
  {Bauerle}}, \bibinfo {author} {\bibfnamefont {J.}~\bibnamefont {Steiner}},
  \bibinfo {author} {\bibfnamefont {L.}~\bibnamefont {Bremicker}}, \bibinfo
  {author} {\bibfnamefont {S.}~\bibnamefont {Schmitt}}, \ and\ \bibinfo
  {author} {\bibfnamefont {C.}~\bibnamefont {Bechinger}},\ }\href
  {http://dx.doi.org/10.1103/PhysRevE.94.012617} {\bibfield  {journal}
  {\bibinfo  {journal} {Phys. Rev. E}\ }\textbf {\bibinfo {volume} {94}},\
  \bibinfo {pages} {012617} (\bibinfo {year} {2016})}\BibitemShut {NoStop}%
\bibitem [{\citenamefont {Lichtner}\ \emph {et~al.}(2012)\citenamefont
  {Lichtner}, \citenamefont {Pototsky},\ and\ \citenamefont
  {Klapp}}]{Lichtner2012}%
  \BibitemOpen
  \bibfield  {author} {\bibinfo {author} {\bibfnamefont {K.}~\bibnamefont
  {Lichtner}}, \bibinfo {author} {\bibfnamefont {A.}~\bibnamefont {Pototsky}},
  \ and\ \bibinfo {author} {\bibfnamefont {S.~H.~L.}\ \bibnamefont {Klapp}},\
  }\href {http://dx.doi.org/10.1103/PhysRevE.86.051405} {\bibfield  {journal}
  {\bibinfo  {journal} {Phys. Rev. E}\ }\textbf {\bibinfo {volume} {86}},\
  \bibinfo {pages} {051405} (\bibinfo {year} {2012})}\BibitemShut {NoStop}%
\bibitem [{\citenamefont {Lichtner}\ and\ \citenamefont
  {Klapp}(2010)}]{Lichtner2010}%
  \BibitemOpen
  \bibfield  {author} {\bibinfo {author} {\bibfnamefont {K.}~\bibnamefont
  {Lichtner}}\ and\ \bibinfo {author} {\bibfnamefont {S.~H.~L.}\ \bibnamefont
  {Klapp}},\ }\href@noop {} {\bibfield  {journal} {\bibinfo  {journal} {EPL}\
  }\textbf {\bibinfo {volume} {92}},\ \bibinfo {pages} {40007} (\bibinfo {year}
  {2010})}\BibitemShut {NoStop}%
\bibitem [{\citenamefont {Hennig}(2009)}]{Hennig2009}%
  \BibitemOpen
  \bibfield  {author} {\bibinfo {author} {\bibfnamefont {D.}~\bibnamefont
  {Hennig}},\ }\href {http://dx.doi.org/10.1103/PhysRevE.79.041114} {\bibfield
  {journal} {\bibinfo  {journal} {Phys. Rev. E}\ }\textbf {\bibinfo {volume}
  {79}},\ \bibinfo {pages} {041114} (\bibinfo {year} {2009})}\BibitemShut
  {NoStop}%
\bibitem [{\citenamefont {Masoller}(2003)}]{Masoller2003}%
  \BibitemOpen
  \bibfield  {author} {\bibinfo {author} {\bibfnamefont {C.}~\bibnamefont
  {Masoller}},\ }\href {http://dx.doi.org/10.1103/PhysRevLett.90.020601}
  {\bibfield  {journal} {\bibinfo  {journal} {Phys. Rev. Lett.}\ }\textbf
  {\bibinfo {volume} {90}},\ \bibinfo {pages} {020601} (\bibinfo {year}
  {2003})}\BibitemShut {NoStop}%
\bibitem [{\citenamefont {Rosinberg}\ \emph {et~al.}(2015)\citenamefont
  {Rosinberg}, \citenamefont {Munakata},\ and\ \citenamefont
  {Tarjus}}]{Rosinberg2015}%
  \BibitemOpen
  \bibfield  {author} {\bibinfo {author} {\bibfnamefont {M.~L.}\ \bibnamefont
  {Rosinberg}}, \bibinfo {author} {\bibfnamefont {T.}~\bibnamefont {Munakata}},
  \ and\ \bibinfo {author} {\bibfnamefont {G.}~\bibnamefont {Tarjus}},\ }{\bibfield  {journal}
  {\bibinfo  {journal} {Phys. Rev. E}\ }\textbf {\bibinfo {volume} {91}},\
  \bibinfo {pages} {042114} (\bibinfo {year} {2015})}\BibitemShut {NoStop}%
\bibitem [{\citenamefont {Guillouzic}\ \emph {et~al.}(1999)\citenamefont
  {Guillouzic}, \citenamefont {L’Heureux},\ and\ \citenamefont
  {Longtin}}]{Guillouzic1999}%
  \BibitemOpen
  \bibfield  {author} {\bibinfo {author} {\bibfnamefont {S.}~\bibnamefont
  {Guillouzic}}, \bibinfo {author} {\bibfnamefont {I.}~\bibnamefont
  {L’Heureux}}, \ and\ \bibinfo {author} {\bibfnamefont {A.}~\bibnamefont
  {Longtin}},\ }\href@noop {} {\bibfield  {journal} {\bibinfo  {journal} {Phys.
  Rev. E}\ }\textbf {\bibinfo {volume} {59}},\ \bibinfo {pages} {3970}
  (\bibinfo {year} {1999})}\BibitemShut {NoStop}%
\bibitem [{\citenamefont {Frank}(2005{\natexlab{a}})}]{Frank2005}%
  \BibitemOpen
  \bibfield  {author} {\bibinfo {author} {\bibfnamefont {T.~D.}\ \bibnamefont
  {Frank}},\ }\href {http://dx.doi.org/10.1103/PhysRevE.71.031106} {\bibfield
  {journal} {\bibinfo  {journal} {Phys. Rev. E}\ }\textbf {\bibinfo {volume}
  {71}},\ \bibinfo {pages} {031106} (\bibinfo {year}
  {2005}{\natexlab{a}})}\BibitemShut {NoStop}%
\bibitem [{\citenamefont {Frank}(2005{\natexlab{b}})}]{Frank2005a}%
  \BibitemOpen
  \bibfield  {author} {\bibinfo {author} {\bibfnamefont {T.~D.}\ \bibnamefont
  {Frank}},\ }\href {http://dx.doi.org/10.1103/PhysRevE.72.011112} {\bibfield
  {journal} {\bibinfo  {journal} {Phys. Rev. E}\ }\textbf {\bibinfo {volume}
  {72}},\ \bibinfo {pages} {011112} (\bibinfo {year}
  {2005}{\natexlab{b}})}\BibitemShut {NoStop}%
\bibitem [{\citenamefont {Giuggioli}\ \emph {et~al.}(2016)\citenamefont
  {Giuggioli}, \citenamefont {McKetterick}, \citenamefont {Kenkre},\ and\
  \citenamefont {Chase}}]{Giuggioli2016}%
  \BibitemOpen
  \bibfield  {author} {\bibinfo {author} {\bibfnamefont {L.}~\bibnamefont
  {Giuggioli}}, \bibinfo {author} {\bibfnamefont {T.~J.}\ \bibnamefont
  {McKetterick}}, \bibinfo {author} {\bibfnamefont {V.~M.}\ \bibnamefont
  {Kenkre}}, \ and\ \bibinfo {author} {\bibfnamefont {M.}~\bibnamefont
  {Chase}},\ }\href {http://stacks.iop.org/1751-8121/49/i=38/a=384002}
  {\bibfield  {journal} {\bibinfo  {journal} {J. Phys. A}\ }\textbf {\bibinfo
  {volume} {49}},\ \bibinfo {pages} {384002} (\bibinfo {year}
  {2016})}\BibitemShut {NoStop}%
\bibitem [{\citenamefont {Seifert}(2012)}]{Seifert2012}%
  \BibitemOpen
  \bibfield  {author} {\bibinfo {author} {\bibfnamefont {U.}~\bibnamefont
  {Seifert}},\ }\href@noop {} {\bibfield  {journal} {\bibinfo  {journal} {Rep.
  Prog. Phys.}\ }\textbf {\bibinfo {volume} {75}},\ \bibinfo {pages} {126001}
  (\bibinfo {year} {2012})}\BibitemShut {NoStop}%
\bibitem [{\citenamefont {Sagawa}\ and\ \citenamefont
  {Ueda}(2012)}]{Sagawa2012}%
  \BibitemOpen
  \bibfield  {author} {\bibinfo {author} {\bibfnamefont {T.}~\bibnamefont
  {Sagawa}}\ and\ \bibinfo {author} {\bibfnamefont {M.}~\bibnamefont {Ueda}},\
  }\href@noop {} {\bibfield  {journal} {\bibinfo  {journal} {Phys. Rev. E}\
  }\textbf {\bibinfo {volume} {85}},\ \bibinfo {pages} {021104} (\bibinfo
  {year} {2012})}\BibitemShut {NoStop}%
\bibitem [{\citenamefont {Abreu}\ and\ \citenamefont
  {Seifert}(2012)}]{Abreu2012}%
  \BibitemOpen
  \bibfield  {author} {\bibinfo {author} {\bibfnamefont {D.}~\bibnamefont
  {Abreu}}\ and\ \bibinfo {author} {\bibfnamefont {U.}~\bibnamefont
  {Seifert}},\ }\href@noop {} {\bibfield  {journal} {\bibinfo  {journal} {Phys.
  Rev. Lett.}\ }\textbf {\bibinfo {volume} {108}},\ \bibinfo {pages} {030601}
  (\bibinfo {year} {2012})}\BibitemShut {NoStop}%
\bibitem [{\citenamefont {Barato}\ and\ \citenamefont
  {Seifert}(2014)}]{Barato2014}%
  \BibitemOpen
  \bibfield  {author} {\bibinfo {author} {\bibfnamefont {A.~C.}\ \bibnamefont
  {Barato}}\ and\ \bibinfo {author} {\bibfnamefont {U.}~\bibnamefont
  {Seifert}},\ }\href {http://dx.doi.org/10.1103/PhysRevLett.112.090601}
  {\bibfield  {journal} {\bibinfo  {journal} {Phys. Rev. Lett.}\ }\textbf
  {\bibinfo {volume} {112}},\ \bibinfo {pages} {090601} (\bibinfo {year}
  {2014})}\BibitemShut {NoStop}%
\bibitem [{\citenamefont {Mandal}\ and\ \citenamefont
  {Jarzynski}(2012)}]{Mandal2012}%
  \BibitemOpen
  \bibfield  {author} {\bibinfo {author} {\bibfnamefont {D.}~\bibnamefont
  {Mandal}}\ and\ \bibinfo {author} {\bibfnamefont {C.}~\bibnamefont
  {Jarzynski}},\ }\href@noop {} {\bibfield  {journal} {\bibinfo  {journal}
  {Proc. Natl. Acad. Sci. U.S.A.}\ }\textbf {\bibinfo {volume} {109}},\
  \bibinfo {pages} {11641} (\bibinfo {year} {2012})}\BibitemShut {NoStop}%
\bibitem [{\citenamefont {Parrondo}\ \emph {et~al.}(2015)\citenamefont
  {Parrondo}, \citenamefont {Horowitz},\ and\ \citenamefont
  {Sagawa}}]{Parrondo2015}%
  \BibitemOpen
  \bibfield  {author} {\bibinfo {author} {\bibfnamefont {J.~M.~R.}\
  \bibnamefont {Parrondo}}, \bibinfo {author} {\bibfnamefont {J.~M.}\
  \bibnamefont {Horowitz}}, \ and\ \bibinfo {author} {\bibfnamefont
  {T.}~\bibnamefont {Sagawa}},\ }\href {http://dx.doi.org/10.1038/nphys3230}
  {\bibfield  {journal} {\bibinfo  {journal} {Nat. Phys.}\ }\textbf {\bibinfo
  {volume} {11}},\ \bibinfo {pages} {131–139} (\bibinfo {year}
  {2015})}\BibitemShut {NoStop}%
\bibitem [{\citenamefont {Kutvonen}\ \emph {et~al.}(2016)\citenamefont
  {Kutvonen}, \citenamefont {Sagawa},\ and\ \citenamefont
  {Ala-Nissila}}]{Kutvonen2016}%
  \BibitemOpen
  \bibfield  {author} {\bibinfo {author} {\bibfnamefont {A.}~\bibnamefont
  {Kutvonen}}, \bibinfo {author} {\bibfnamefont {T.}~\bibnamefont {Sagawa}}, \
  and\ \bibinfo {author} {\bibfnamefont {T.}~\bibnamefont {Ala-Nissila}},\
  }\href {http://dx.doi.org/10.1103/PhysRevE.93.032147} {\bibfield  {journal}
  {\bibinfo  {journal} {Phys. Rev. E}\ }\textbf {\bibinfo {volume} {93}},\
  \bibinfo {pages} {032147} (\bibinfo {year} {2016})}\BibitemShut {NoStop}%
\bibitem [{\citenamefont {Loos}\ \emph {et~al.}(2014)\citenamefont {Loos},
  \citenamefont {Gernert},\ and\ \citenamefont {Klapp}}]{Loos2014}%
  \BibitemOpen
  \bibfield  {author} {\bibinfo {author} {\bibfnamefont {S.~A.~M.}\
  \bibnamefont {Loos}}, \bibinfo {author} {\bibfnamefont {R.}~\bibnamefont
  {Gernert}}, \ and\ \bibinfo {author} {\bibfnamefont {S.~H.~L.}\ \bibnamefont
  {Klapp}},\ }\href {http://dx.doi.org/10.1103/PhysRevE.89.052136} {\bibfield
  {journal} {\bibinfo  {journal} {Phys. Rev. E}\ }\textbf {\bibinfo {volume}
  {89}},\ \bibinfo {pages} {052136} (\bibinfo {year} {2014})}\BibitemShut
  {NoStop}%
\bibitem [{\citenamefont {Kim}\ and\ \citenamefont {Qian}(2007)}]{Kim2007}%
  \BibitemOpen
  \bibfield  {author} {\bibinfo {author} {\bibfnamefont {K.~H.}\ \bibnamefont
  {Kim}}\ and\ \bibinfo {author} {\bibfnamefont {H.}~\bibnamefont {Qian}},\
  }\href {http://dx.doi.org/10.1103/PhysRevE.75.022102} {\bibfield  {journal}
  {\bibinfo  {journal} {Phys. Rev. E}\ }\textbf {\bibinfo {volume} {75}},\
  \bibinfo {pages} {022102} (\bibinfo {year} {2007})}\BibitemShut {NoStop}%
\bibitem [{\citenamefont {Koski}\ \emph {et~al.}(2014)\citenamefont {Koski},
  \citenamefont {Maisi}, \citenamefont {Sagawa},\ and\ \citenamefont
  {Pekola}}]{Koski2014}%
  \BibitemOpen
  \bibfield  {author} {\bibinfo {author} {\bibfnamefont {J.~V.}\ \bibnamefont
  {Koski}}, \bibinfo {author} {\bibfnamefont {V.~F.}\ \bibnamefont {Maisi}},
  \bibinfo {author} {\bibfnamefont {T.}~\bibnamefont {Sagawa}}, \ and\ \bibinfo
  {author} {\bibfnamefont {J.~P.}\ \bibnamefont {Pekola}},\ }\href
  {http://dx.doi.org/10.1103/PhysRevLett.113.030601} {\bibfield  {journal}
  {\bibinfo  {journal} {Phys. Rev. Lett.}\ }\textbf {\bibinfo {volume} {113}},\
  \bibinfo {pages} {030601} (\bibinfo {year} {2014})}\BibitemShut {NoStop}%
\bibitem [{\citenamefont {Jiang}\ \emph {et~al.}(2011)\citenamefont {Jiang},
  \citenamefont {Xiao},\ and\ \citenamefont {Hou}}]{Jiang2011}%
  \BibitemOpen
  \bibfield  {author} {\bibinfo {author} {\bibfnamefont {H.}~\bibnamefont
  {Jiang}}, \bibinfo {author} {\bibfnamefont {T.}~\bibnamefont {Xiao}}, \ and\
  \bibinfo {author} {\bibfnamefont {Z.}~\bibnamefont {Hou}},\ }\href
  {http://dx.doi.org/10.1103/PhysRevE.83.061144} {\bibfield  {journal}
  {\bibinfo  {journal} {Phys. Rev. E}\ }\textbf {\bibinfo {volume} {83}},\
  \bibinfo {pages} {061144} (\bibinfo {year} {2011})}\BibitemShut {NoStop}%
\bibitem [{\citenamefont {Munakata}\ \emph {et~al.}(2009)\citenamefont
  {Munakata}, \citenamefont {Iwama},\ and\ \citenamefont
  {Kimizuka}}]{Munakata2009}%
  \BibitemOpen
  \bibfield  {author} {\bibinfo {author} {\bibfnamefont {T.}~\bibnamefont
  {Munakata}}, \bibinfo {author} {\bibfnamefont {S.}~\bibnamefont {Iwama}}, \
  and\ \bibinfo {author} {\bibfnamefont {M.}~\bibnamefont {Kimizuka}},\ }\href
  {http://dx.doi.org/10.1103/PhysRevE.79.031104} {\bibfield  {journal}
  {\bibinfo  {journal} {Phys. Rev. E}\ }\textbf {\bibinfo {volume} {79}},\
  \bibinfo {pages} {031104} (\bibinfo {year} {2009})}\BibitemShut {NoStop}%
\bibitem [{\citenamefont {Munakata}\ and\ \citenamefont
  {Rosinberg}(2014)}]{Munakata2014}%
  \BibitemOpen
  \bibfield  {author} {\bibinfo {author} {\bibfnamefont {T.}~\bibnamefont
  {Munakata}}\ and\ \bibinfo {author} {\bibfnamefont {M.~L.}\ \bibnamefont
  {Rosinberg}},\ }\href@noop {} {\bibfield  {journal} {\bibinfo  {journal}
  {Phys. Rev. Lett.}\ }\textbf {\bibinfo {volume} {112}},\ \bibinfo {pages}
  {180601} (\bibinfo {year} {2014})}\BibitemShut {NoStop}%
\bibitem [{\citenamefont {Küchler}\ and\ \citenamefont
  {Mensch}(1992)}]{Kuechler1992}%
  \BibitemOpen
  \bibfield  {author} {\bibinfo {author} {\bibfnamefont {U.}~\bibnamefont
  {Küchler}}\ and\ \bibinfo {author} {\bibfnamefont {B.}~\bibnamefont
  {Mensch}},\ }\href {http://dx.doi.org/10.1080/17442509208833780} {\bibfield
  {journal} {\bibinfo  {journal} {Stoch. Stoch. Rep.}\ }\textbf {\bibinfo
  {volume} {40}},\ \bibinfo {pages} {23–42} (\bibinfo {year}
  {1992})}\BibitemShut {NoStop}%
\bibitem [{\citenamefont {Frank}\ and\ \citenamefont {Beek}(2001)}]{Frank2001}%
  \BibitemOpen
  \bibfield  {author} {\bibinfo {author} {\bibfnamefont {T.~D.}\ \bibnamefont
  {Frank}}\ and\ \bibinfo {author} {\bibfnamefont {P.~J.}\ \bibnamefont
  {Beek}},\ }\href {http://dx.doi.org/10.1103/PhysRevE.64.021917} {\bibfield
  {journal} {\bibinfo  {journal} {Phys. Rev. E}\ }\textbf {\bibinfo {volume}
  {64}},\ \bibinfo {pages} {021917} (\bibinfo {year} {2001})}\BibitemShut
  {NoStop}%
\bibitem [{\citenamefont {Guillouzic}\ \emph {et~al.}(2000)\citenamefont
  {Guillouzic}, \citenamefont {L’Heureux},\ and\ \citenamefont
  {Longtin}}]{Guillouzic2000}%
  \BibitemOpen
  \bibfield  {author} {\bibinfo {author} {\bibfnamefont {S.}~\bibnamefont
  {Guillouzic}}, \bibinfo {author} {\bibfnamefont {I.}~\bibnamefont
  {L’Heureux}}, \ and\ \bibinfo {author} {\bibfnamefont {A.}~\bibnamefont
  {Longtin}},\ }\href@noop {} {\bibfield  {journal} {\bibinfo  {journal} {Phys.
  Rev. E}\ }\textbf {\bibinfo {volume} {61}},\ \bibinfo {pages} {4906}
  (\bibinfo {year} {2000})}\BibitemShut {NoStop}%
\bibitem [{\citenamefont {Marconi}\ and\ \citenamefont
  {Tarazona}(1999)}]{Marconi1999}%
  \BibitemOpen
  \bibfield  {author} {\bibinfo {author} {\bibfnamefont {U.~M.~B.}\
  \bibnamefont {Marconi}}\ and\ \bibinfo {author} {\bibfnamefont
  {P.}~\bibnamefont {Tarazona}},\ }\href {\doibase 10.1063/1.478705} {\bibfield
   {journal} {\bibinfo  {journal} {J. Chem. Phys.}\ }\textbf
  {\bibinfo {volume} {110}},\ \bibinfo {pages} {8032} (\bibinfo {year}
  {1999})} \BibitemShut {NoStop}%
\bibitem [{\citenamefont {Menzel}\ \emph {et~al.}(2016)\citenamefont {Menzel},
    \citenamefont {Saha}, \citenamefont {Hoell},\ and\ \citenamefont
    {L{\"o}wen}}]{Menzel2016}%
    \BibitemOpen
    \bibfield  {author} {\bibinfo {author} {\bibfnamefont {A.~M.}\ \bibnamefont
    {Menzel}}, \bibinfo {author} {\bibfnamefont {A.}~\bibnamefont {Saha}},
    \bibinfo {author} {\bibfnamefont {C.}~\bibnamefont {Hoell}}, \ and\ \bibinfo
    {author} {\bibfnamefont {H.}~\bibnamefont {L{\"o}wen}},\ }\href@noop {}
    {\bibfield  {journal} {\bibinfo  {journal} {J. Chem. Phys.}\
    }\textbf {\bibinfo {volume} {144}},\ \bibinfo {pages} {024115} (\bibinfo
    {year} {2016})}\BibitemShut {NoStop}%
\bibitem [{\citenamefont {Das}(2004)}]{Das2004}%
  \BibitemOpen
  \bibfield  {author} {\bibinfo {author} {\bibfnamefont {S.~P.}\ \bibnamefont
  {Das}},\ }\href@noop {} {\bibfield  {journal} {\bibinfo  {journal} {Rev. Mod. Phys.}\ }\textbf {\bibinfo {volume} {76}},\ \bibinfo {pages}
  {785} (\bibinfo {year} {2004})}\BibitemShut {NoStop}%
\bibitem [{\citenamefont {Janssen}\ and\ \citenamefont
  {Reichman}(2015)}]{Janssen2015}%
  \BibitemOpen
  \bibfield  {author} {\bibinfo {author} {\bibfnamefont {L.~M.~C.}\ \bibnamefont
  {Janssen}}\ and\ \bibinfo {author} {\bibfnamefont {D.~R.}\ \bibnamefont
  {Reichman}},\ }\href@noop {} {\bibfield  {journal} {\bibinfo  {journal}
  {Phys. Rev. Lett.}\ }\textbf {\bibinfo {volume} {115}},\ \bibinfo
  {pages} {205701} (\bibinfo {year} {2015})}\BibitemShut {NoStop}%
\bibitem [{\citenamefont {Kotar}\ \emph {et~al.}(2010)\citenamefont {Kotar},
  \citenamefont {Leoni}, \citenamefont {Bassetti}, \citenamefont
  {Lagomarsino},\ and\ \citenamefont {Cicuta}}]{Kotar2010}%
  \BibitemOpen
  \bibfield  {author} {\bibinfo {author} {\bibfnamefont {J.}~\bibnamefont
  {Kotar}}, \bibinfo {author} {\bibfnamefont {M.}~\bibnamefont {Leoni}},
  \bibinfo {author} {\bibfnamefont {B.}~\bibnamefont {Bassetti}}, \bibinfo
  {author} {\bibfnamefont {M.~C.}\ \bibnamefont {Lagomarsino}}, \ and\ \bibinfo
  {author} {\bibfnamefont {P.}~\bibnamefont {Cicuta}},\ }\href@noop {}
  {\bibfield  {journal} {\bibinfo  {journal} {Proc. Natl. Acad. Sci. U.S.A.}\
  }\textbf {\bibinfo {volume} {107}},\ \bibinfo {pages} {7669} (\bibinfo {year}
  {2010})}\BibitemShut {NoStop}%
\bibitem [{\citenamefont {Qian}\ \emph {et~al.}(2013)\citenamefont {Qian},
  \citenamefont {Montiel}, \citenamefont {Bregulla}, \citenamefont {Cichos},\
  and\ \citenamefont {Yang}}]{Qian2013}%
  \BibitemOpen
  \bibfield  {author} {\bibinfo {author} {\bibfnamefont {B.}~\bibnamefont
  {Qian}}, \bibinfo {author} {\bibfnamefont {D.}~\bibnamefont {Montiel}},
  \bibinfo {author} {\bibfnamefont {A.}~\bibnamefont {Bregulla}}, \bibinfo
  {author} {\bibfnamefont {F.}~\bibnamefont {Cichos}}, \ and\ \bibinfo {author}
  {\bibfnamefont {H.}~\bibnamefont {Yang}},\ }\href@noop {} {\bibfield
  {journal} {\bibinfo  {journal} {Chem. Sci.}\ }\textbf {\bibinfo {volume}
  {4}},\ \bibinfo {pages} {1420} (\bibinfo {year} {2013})}\BibitemShut
  {NoStop}%
\bibitem [{\citenamefont {Balijepalli}\ \emph {et~al.}(2012)\citenamefont
  {Balijepalli}, \citenamefont {Gorman}, \citenamefont {Gupta},\ and\
  \citenamefont {LeBrun}}]{Balijepalli2012}%
  \BibitemOpen
  \bibfield  {author} {\bibinfo {author} {\bibfnamefont {A.}~\bibnamefont
  {Balijepalli}}, \bibinfo {author} {\bibfnamefont {J.~J.}\ \bibnamefont
  {Gorman}}, \bibinfo {author} {\bibfnamefont {S.~K.}\ \bibnamefont {Gupta}}, \
  and\ \bibinfo {author} {\bibfnamefont {T.~W.}\ \bibnamefont {LeBrun}},\
  }\href@noop {} {\bibfield  {journal} {\bibinfo  {journal} {Nano Lett.}\
  }\textbf {\bibinfo {volume} {12}},\ \bibinfo {pages} {2347} (\bibinfo {year}
  {2012})}\BibitemShut {NoStop}%
\bibitem [{\citenamefont {Gernert}\ \emph {et~al.}(2014)\citenamefont
  {Gernert}, \citenamefont {Emary},\ and\ \citenamefont {Klapp}}]{Gernert2014}%
  \BibitemOpen
  \bibfield  {author} {\bibinfo {author} {\bibfnamefont {R.}~\bibnamefont
  {Gernert}}, \bibinfo {author} {\bibfnamefont {C.}~\bibnamefont {Emary}}, \
  and\ \bibinfo {author} {\bibfnamefont {S.~H.~L.}\ \bibnamefont {Klapp}},\
  }\href@noop {} {\bibfield  {journal} {\bibinfo  {journal} {Phys. Rev. E}\
  }\textbf {\bibinfo {volume} {90}},\ \bibinfo {pages} {062115} (\bibinfo
  {year} {2014})}\BibitemShut {NoStop}%
\bibitem [{\citenamefont {Gernert}\ and\ \citenamefont
  {Klapp}(2015)}]{Gernert2015}%
  \BibitemOpen
  \bibfield  {author} {\bibinfo {author} {\bibfnamefont {R.}~\bibnamefont
  {Gernert}}\ and\ \bibinfo {author} {\bibfnamefont {S.~H.~L.}\ \bibnamefont
  {Klapp}},\ }\href@noop {} {\bibfield  {journal} {\bibinfo  {journal} {Phys.
  Rev. E}\ }\textbf {\bibinfo {volume} {92}},\ \bibinfo {pages} {022132}
  (\bibinfo {year} {2015})}\BibitemShut {NoStop}%
\bibitem [{\citenamefont {W{\"o}rdemann}(2012)}]{Woerdemann2012}%
  \BibitemOpen
  \bibfield  {author} {\bibinfo {author} {\bibfnamefont {M.}~\bibnamefont
  {W{\"o}rdemann}},\ }\href@noop {} {\emph {\bibinfo {title} {Structured Light
  Fields: Applications in Optical Trapping, Manipulation, and Organisation}}}\
  (\bibinfo  {publisher} {Springer Science \& Business Media},\ \bibinfo {year}
  {2012})\BibitemShut {NoStop}%
\bibitem [{\citenamefont {Reimann}\ \emph {et~al.}(2001)\citenamefont
  {Reimann}, \citenamefont {Van~den Broeck}, \citenamefont {Linke},
  \citenamefont {H{\"a}nggi}, \citenamefont {Rubi},\ and\ \citenamefont
  {P{\'e}rez-Madrid}}]{Reimann2001}%
  \BibitemOpen
  \bibfield  {author} {\bibinfo {author} {\bibfnamefont {P.}~\bibnamefont
  {Reimann}}, \bibinfo {author} {\bibfnamefont {C.}~\bibnamefont {Van~den
  Broeck}}, \bibinfo {author} {\bibfnamefont {H.}~\bibnamefont {Linke}},
  \bibinfo {author} {\bibfnamefont {P.}~\bibnamefont {H{\"a}nggi}}, \bibinfo
  {author} {\bibfnamefont {J.~M.}\ \bibnamefont {Rubi}}, \ and\ \bibinfo
  {author} {\bibfnamefont {A.}~\bibnamefont {P{\'e}rez-Madrid}},\ }\href
  {http://dx.doi.org/10.1103/PhysRevLett.87.010602} {\bibfield  {journal}
  {\bibinfo  {journal} {Phys. Rev. Lett.}\ }\textbf {\bibinfo {volume} {87}},\
  \bibinfo {pages} {010602} (\bibinfo {year} {2001})}\BibitemShut {NoStop}%
\bibitem [{\citenamefont {Juniper}\ \emph {et~al.}(2016)\citenamefont
  {Juniper}, \citenamefont {Straube}, \citenamefont {Aarts},\ and\
  \citenamefont {Dullens}}]{Juniper2016}%
  \BibitemOpen
  \bibfield  {author} {\bibinfo {author} {\bibfnamefont {M.~P.~N.}\
  \bibnamefont {Juniper}}, \bibinfo {author} {\bibfnamefont {A.~V.}\
  \bibnamefont {Straube}}, \bibinfo {author} {\bibfnamefont {D.~G. A.~L.}\
  \bibnamefont {Aarts}}, \ and\ \bibinfo {author} {\bibfnamefont {R.~P.~A.}\
  \bibnamefont {Dullens}},\ }\href
  {http://dx.doi.org/10.1103/PhysRevE.93.012608} {\bibfield  {journal}
  {\bibinfo  {journal} {Phys. Rev. E}\ }\textbf {\bibinfo {volume} {93}},\
  \bibinfo {pages} {012608} (\bibinfo {year} {2016})}\BibitemShut {NoStop}%
\bibitem [{\citenamefont {Emary}\ \emph {et~al.}(2012)\citenamefont {Emary},
  \citenamefont {Gernert},\ and\ \citenamefont {Klapp}}]{Emary2012}%
  \BibitemOpen
  \bibfield  {author} {\bibinfo {author} {\bibfnamefont {C.}~\bibnamefont
  {Emary}}, \bibinfo {author} {\bibfnamefont {R.}~\bibnamefont {Gernert}}, \
  and\ \bibinfo {author} {\bibfnamefont {S.~H.~L.}\ \bibnamefont {Klapp}},\
  }\href@noop {} {\bibfield  {journal} {\bibinfo  {journal} {Phys. Rev. E}\
  }\textbf {\bibinfo {volume} {86}},\ \bibinfo {pages} {061135} (\bibinfo
  {year} {2012})}\BibitemShut {NoStop}%
\bibitem [{\citenamefont {Kramers}(1940)}]{Kramers1940}%
  \BibitemOpen
  \bibfield  {author} {\bibinfo {author} {\bibfnamefont {H.~A.}\ \bibnamefont
  {Kramers}},\ }\href@noop {} {\bibfield  {journal} {\bibinfo  {journal}
  {Physica}\ }\textbf {\bibinfo {volume} {7}},\ \bibinfo {pages} {284}
  (\bibinfo {year} {1940})}\BibitemShut {NoStop}%
\bibitem [{\citenamefont {H{\"a}nggi}\ \emph {et~al.}(1990)\citenamefont
  {H{\"a}nggi}, \citenamefont {Talkner},\ and\ \citenamefont
  {Borkovec}}]{Hanggi1990}%
  \BibitemOpen
  \bibfield  {author} {\bibinfo {author} {\bibfnamefont {P.}~\bibnamefont
  {H{\"a}nggi}}, \bibinfo {author} {\bibfnamefont {P.}~\bibnamefont {Talkner}},
  \ and\ \bibinfo {author} {\bibfnamefont {M.}~\bibnamefont {Borkovec}},\
  }\href@noop {} {\bibfield  {journal} {\bibinfo  {journal} {Rev. Mod. Phys.}\
  }\textbf {\bibinfo {volume} {62}},\ \bibinfo {pages} {251} (\bibinfo {year}
  {1990})}\BibitemShut {NoStop}%
\bibitem [{\citenamefont {Risken}(1984)}]{Risken1984}%
  \BibitemOpen
  \bibfield  {author} {\bibinfo {author} {\bibfnamefont {H.}~\bibnamefont
  {Risken}},\ }\href@noop {} {\emph {\bibinfo {title} {The Fokker-Planck
  Equation}}}\ (\bibinfo  {publisher} {Springer},\ \bibinfo {year}
  {1984})\BibitemShut {NoStop}%
\bibitem [{\citenamefont {Gardiner}(2002)}]{Gardiner2002}%
  \BibitemOpen
  \bibfield  {author} {\bibinfo {author} {\bibfnamefont {C.~W.}\ \bibnamefont
  {Gardiner}},\ }\href@noop {} {\emph {\bibinfo {title} {Handbook of Stochastic
  Methods}}},\ \bibinfo {edition} {2nd}\ ed.,\ \bibinfo {number} {6th
  printing}\ (\bibinfo  {publisher} {Springer, Berlin--Heidelberg--New York},\
  \bibinfo {year} {2002})\BibitemShut {NoStop}%
\bibitem [{\citenamefont {Goulding}\ \emph {et~al.}(2007)\citenamefont
  {Goulding}, \citenamefont {Melnik}, \citenamefont {Curtin}, \citenamefont
  {Piwonski}, \citenamefont {Houlihan}, \citenamefont {Gleeson},\ and\
  \citenamefont {Huyet}}]{Goulding2007}%
  \BibitemOpen
  \bibfield  {author} {\bibinfo {author} {\bibfnamefont {D.}~\bibnamefont
  {Goulding}}, \bibinfo {author} {\bibfnamefont {S.}~\bibnamefont {Melnik}},
  \bibinfo {author} {\bibfnamefont {D.}~\bibnamefont {Curtin}}, \bibinfo
  {author} {\bibfnamefont {T.}~\bibnamefont {Piwonski}}, \bibinfo {author}
  {\bibfnamefont {J.}~\bibnamefont {Houlihan}}, \bibinfo {author}
  {\bibfnamefont {J.~P.}\ \bibnamefont {Gleeson}}, \ and\ \bibinfo {author}
  {\bibfnamefont {G.}~\bibnamefont {Huyet}},\ }\href@noop {} {\bibfield
  {journal} {\bibinfo  {journal} {Phys. Rev. E}\ }\textbf {\bibinfo {volume}
  {76}},\ \bibinfo {pages} {031128} (\bibinfo {year} {2007})}\BibitemShut
  {NoStop}%
\bibitem [{\citenamefont {Tsimring}\ and\ \citenamefont
  {Pikovsky}(2001)}]{Tsimring2001}%
  \BibitemOpen
  \bibfield  {author} {\bibinfo {author} {\bibfnamefont {L.~S.}\ \bibnamefont
  {Tsimring}}\ and\ \bibinfo {author} {\bibfnamefont {A.}~\bibnamefont
  {Pikovsky}},\ }\href {http://dx.doi.org/10.1103/PhysRevLett.87.250602}
  {\bibfield  {journal} {\bibinfo  {journal} {Phys. Rev. Lett.}\ }\textbf
  {\bibinfo {volume} {87}},\ \bibinfo {pages} {250602} (\bibinfo {year}
  {2001})}\BibitemShut {NoStop}%
\bibitem [{\citenamefont {Du}\ and\ \citenamefont {Mei}(2015)}]{Du2015}%
  \BibitemOpen
  \bibfield  {author} {\bibinfo {author} {\bibfnamefont {L.}~\bibnamefont
  {Du}}\ and\ \bibinfo {author} {\bibfnamefont {D.}~\bibnamefont {Mei}},\
  }\href@noop {} {\bibfield  {journal} {\bibinfo  {journal} {Indian J. Phys.}\
  }\textbf {\bibinfo {volume} {89}},\ \bibinfo {pages} {267} (\bibinfo {year}
  {2015})}\BibitemShut {NoStop}%
\bibitem [{\citenamefont {Xiao}(2016)}]{Xiao2016}%
  \BibitemOpen
  \bibfield  {author} {\bibinfo {author} {\bibfnamefont {T.}~\bibnamefont
  {Xiao}},\ }\href@noop {} {\bibfield  {journal} {\bibinfo  {journal} {Phys.
  Rev. E}\ }\textbf {\bibinfo {volume} {94}},\ \bibinfo {pages} {052109}
  (\bibinfo {year} {2016})}\BibitemShut {NoStop}%
\bibitem [{\citenamefont {Piwonski}\ \emph {et~al.}(2005)\citenamefont
  {Piwonski}, \citenamefont {Houlihan}, \citenamefont {Busch},\ and\
  \citenamefont {Huyet}}]{Piwonski2005}%
  \BibitemOpen
  \bibfield  {author} {\bibinfo {author} {\bibfnamefont {T.}~\bibnamefont
  {Piwonski}}, \bibinfo {author} {\bibfnamefont {J.}~\bibnamefont {Houlihan}},
  \bibinfo {author} {\bibfnamefont {T.}~\bibnamefont {Busch}}, \ and\ \bibinfo
  {author} {\bibfnamefont {G.}~\bibnamefont {Huyet}},\ }\href
  {http://dx.doi.org/10.1103/PhysRevLett.95.040601} {\bibfield  {journal}
  {\bibinfo  {journal} {Phys. Rev. Lett.}\ }\textbf {\bibinfo {volume} {95}},\
  \bibinfo {pages} {040601} (\bibinfo {year} {2005})}\BibitemShut {NoStop}%
\bibitem [{\citenamefont {Novikov}(1965)}]{Novikov1965}%
  \BibitemOpen
  \bibfield  {author} {\bibinfo {author} {\bibfnamefont {E.~A.}\ \bibnamefont
  {Novikov}},\ }\href@noop {} {\bibfield  {journal} {\bibinfo  {journal} {Sov.
  Phys. JETP}\ }\textbf {\bibinfo {volume} {20}},\ \bibinfo {pages} {1290}
  (\bibinfo {year} {1965})}\BibitemShut {NoStop}%
\bibitem [{\citenamefont {Frank}\ \emph {et~al.}(2003)\citenamefont {Frank},
  \citenamefont {Beek},\ and\ \citenamefont {Friedrich}}]{Frank2003}%
  \BibitemOpen
  \bibfield  {author} {\bibinfo {author} {\bibfnamefont {T.~D.}\ \bibnamefont
  {Frank}}, \bibinfo {author} {\bibfnamefont {P.~J.}\ \bibnamefont {Beek}}, \
  and\ \bibinfo {author} {\bibfnamefont {R.}~\bibnamefont {Friedrich}},\ }\href
  {http://dx.doi.org/10.1103/PhysRevE.68.021912} {\bibfield  {journal}
  {\bibinfo  {journal} {Phys. Rev. E}\ }\textbf {\bibinfo {volume} {68}},\
  \bibinfo {pages} {021912} (\bibinfo {year} {2003})}\BibitemShut {NoStop}%
\bibitem [{\citenamefont {Pyragas}(1992)}]{Pyragas1992}%
  \BibitemOpen
  \bibfield  {author} {\bibinfo {author} {\bibfnamefont {K.}~\bibnamefont
  {Pyragas}},\ }\href@noop {} {\bibfield  {journal} {\bibinfo  {journal} {Phys.
  Lett. A}\ }\textbf {\bibinfo {volume} {170}},\ \bibinfo {pages} {421}
  (\bibinfo {year} {1992})}\BibitemShut {NoStop}%
\bibitem [{\citenamefont {Nbendjo}\ \emph {et~al.}(2003)\citenamefont
  {Nbendjo}, \citenamefont {Tchoukuegno},\ and\ \citenamefont
  {Woafo}}]{Nbendjo2003}%
  \BibitemOpen
  \bibfield  {author} {\bibinfo {author} {\bibfnamefont {B.~N.}\ \bibnamefont
  {Nbendjo}}, \bibinfo {author} {\bibfnamefont {R.}~\bibnamefont
  {Tchoukuegno}}, \ and\ \bibinfo {author} {\bibfnamefont {P.}~\bibnamefont
  {Woafo}},\ }\href@noop {} {\bibfield  {journal} {\bibinfo  {journal} {Chaos,
  Solitons \& Fractals}\ }\textbf {\bibinfo {volume} {18}},\ \bibinfo {pages}
  {345} (\bibinfo {year} {2003})}\BibitemShut {NoStop}%
\bibitem [{\citenamefont {McKetterick}\ and\ \citenamefont
  {Giuggioli}(2014)}]{McKetterick2014}%
  \BibitemOpen
  \bibfield  {author} {\bibinfo {author} {\bibfnamefont {T.~J.}\ \bibnamefont
  {McKetterick}}\ and\ \bibinfo {author} {\bibfnamefont {L.}~\bibnamefont
  {Giuggioli}},\ }\href {\doibase 10.1103/PhysRevE.90.042135} {\bibfield
  {journal} {\bibinfo  {journal} {Phys. Rev. E}\ }\textbf {\bibinfo {volume}
  {90}},\ \bibinfo {pages} {042135} (\bibinfo {year} {2014})}\BibitemShut
  {NoStop}%
\bibitem [{\citenamefont {H{\"a}nggi}\ and\ \citenamefont
  {Talkner}(1978)}]{Hanggi1978}%
  \BibitemOpen
  \bibfield  {author} {\bibinfo {author} {\bibfnamefont {P.}~\bibnamefont
  {H{\"a}nggi}}\ and\ \bibinfo {author} {\bibfnamefont {P.}~\bibnamefont
  {Talkner}},\ }\href@noop {} {\bibfield  {journal} {\bibinfo  {journal} {Phys.
  Lett. A}\ }\textbf {\bibinfo {volume} {68}},\ \bibinfo {pages} {9} (\bibinfo
  {year} {1978})}\BibitemShut {NoStop}%
\bibitem [{\citenamefont {H{\"a}nggi}\ and\ \citenamefont
  {Thomas}(1982)}]{Hanggi1982}%
  \BibitemOpen
  \bibfield  {author} {\bibinfo {author} {\bibfnamefont {P.}~\bibnamefont
  {H{\"a}nggi}}\ and\ \bibinfo {author} {\bibfnamefont {H.}~\bibnamefont
  {Thomas}},\ }\href@noop {} {\bibfield  {journal} {\bibinfo  {journal} {Phys.
  Rep.}\ }\textbf {\bibinfo {volume} {88}},\ \bibinfo {pages} {207} (\bibinfo
  {year} {1982})}\BibitemShut {NoStop}%
\bibitem [{\citenamefont {Hern{\'a}ndez-Machado}\ \emph
  {et~al.}(1983)\citenamefont {Hern{\'a}ndez-Machado}, \citenamefont {Sancho},
  \citenamefont {San~Miguel},\ and\ \citenamefont {Pesquera}}]{Hernandez1983}%
  \BibitemOpen
  \bibfield  {author} {\bibinfo {author} {\bibfnamefont {A.}~\bibnamefont
  {Hern{\'a}ndez-Machado}}, \bibinfo {author} {\bibfnamefont {J.}~\bibnamefont
  {Sancho}}, \bibinfo {author} {\bibfnamefont {M.}~\bibnamefont {San~Miguel}},
  \ and\ \bibinfo {author} {\bibfnamefont {L.}~\bibnamefont {Pesquera}},\
  }\href@noop {} {\bibfield  {journal} {\bibinfo  {journal} {EPJ B}\ }\textbf
  {\bibinfo {volume} {52}},\ \bibinfo {pages} {335} (\bibinfo {year}
  {1983})}\BibitemShut {NoStop}%
\bibitem [{\citenamefont {Budini}\ and\ \citenamefont
  {C{\'a}ceres}(2004)}]{Budini2004}%
  \BibitemOpen
  \bibfield  {author} {\bibinfo {author} {\bibfnamefont {A.~A.}\ \bibnamefont
  {Budini}}\ and\ \bibinfo {author} {\bibfnamefont {M.~O.}\ \bibnamefont
  {C{\'a}ceres}},\ }\href@noop {} {\bibfield  {journal} {\bibinfo  {journal}
  {J. Phys. A}\ }\textbf {\bibinfo {volume} {37}},\ \bibinfo {pages} {5959}
  (\bibinfo {year} {2004})}\BibitemShut {NoStop}%
\bibitem [{\citenamefont {Frank}(2004)}]{Frank2004}%
  \BibitemOpen
  \bibfield  {author} {\bibinfo {author} {\bibfnamefont {T.~D.}\ \bibnamefont
  {Frank}},\ }\href {http://dx.doi.org/10.1103/PhysRevE.69.061104} {\bibfield
  {journal} {\bibinfo  {journal} {Phys. Rev. E}\ }\textbf {\bibinfo {volume}
  {69}},\ \bibinfo {pages} {061104} (\bibinfo {year} {2004})}\BibitemShut
  {NoStop}%
\bibitem [{\citenamefont {Buckwar}(2000)}]{Buckwar2000}%
  \BibitemOpen
  \bibfield  {author} {\bibinfo {author} {\bibfnamefont {E.}~\bibnamefont
  {Buckwar}},\ }\href@noop {} {\bibfield  {journal} {\bibinfo  {journal} {J.
  Comput. Appl. Math.}\ }\textbf {\bibinfo {volume} {125}},\ \bibinfo {pages}
  {297} (\bibinfo {year} {2000})}\BibitemShut {NoStop}%
\bibitem [{\citenamefont {Kloeden}\ and\ \citenamefont
  {Platen}(1992)}]{Kloeden1992}%
  \BibitemOpen
  \bibfield  {author} {\bibinfo {author} {\bibfnamefont {P.}~\bibnamefont
  {Kloeden}}\ and\ \bibinfo {author} {\bibfnamefont {E.}~\bibnamefont
  {Platen}},\ }\href@noop {} {\emph {\bibinfo {title} {Numerical solution of
  stochastic differential equations}}}\ (\bibinfo  {publisher}
  {Springer--Verlag},\ \bibinfo {year} {1992})\BibitemShut {NoStop}%
\bibitem [{Mer(2016)}]{Mersenne}%
  \BibitemOpen
  \href@noop {} {} (\bibinfo {year} {June 2016}),\ \bibinfo {note} {weblink:
  \url{www.math.sci.hiroshima-u.ac.jp/~m-mat/MT/emt.html}}\BibitemShut
  {NoStop}%
\end{thebibliography}
\end{document}